\documentclass[%
preprint,
 amsmath,amssymb,
 aps,
pra,
floatfix,
]{revtex4-2}

\usepackage{graphicx}
\usepackage{dcolumn}
\usepackage{bm}
\usepackage{hyperref}

\usepackage{bbding}
\usepackage{braket}
\usepackage{graphicx}
\usepackage{multirow}

\usepackage{epstopdf,cancel,ulem}
\usepackage{epsf,latexsym,bbm,euscript}
\usepackage{hyperref}

\usepackage{tcolorbox}

\usepackage{color}

\newcommand{\blue}[1]{\textcolor{blue}{#1}}

\begin{document}

\title{A Kerr kernel  quantum learning machine. }

 \author{Carolyn Wood, Sally Shrapnel, G J Milburn}
 \email{g.milburn@uq.edu.au}
\affiliation{Centre for Engineered Quantum Systems, School of Mathematics and Physics, The University of Queensland.}

\date{\today}

\begin{abstract}
Kernel methods are of current interest in quantum machine learning due to similarities with quantum computing in how they process information in high-dimensional feature (Hilbert) spaces. Kernels are believed to offer particular advantages when they cannot be computed classically, so a kernel matrix with indisputably nonclassical elements is desirable provided it can be generated efficiently in a particular physical machine.  Kerr nonlinearities, known to be a route to universal continuous variable (CV) quantum computation, may be able to play this role for quantum machine learning. We propose a quantum hardware kernel implementation scheme based on superconducting quantum circuits. The scheme does not use qubits or quantum circuits but rather exploits the analogue features of Kerr coupled modes. Our approach is more akin to the growing number of analog machine learning schemes based on sampling quantum probabilities directly in an engineered device by stochastic quantum control.
\end{abstract}

\maketitle


\section{Introduction}

In this paper we propose a scheme for a quantum kernel machine learning protocol based on quantum Kerr non linearities for bosonic modes. This is a fundamentally analogue scheme. Unlike much of quantum machine learning, our scheme is not based on the quantum computer paradigm with qubits and circuits of single and two qubit gates\cite{garcia2023}. Our approach is more akin to the growing number of analog machine learning schemes based on sampling quantum probabilities in a device engineered directly\cite{kendall2020training,wright,wang2024training,hard-is-soft,coles2023thermodynamic,extropic}. In our scheme, kernel matrices are not computed but directly sampled through measurements made on the device.  They can then be passed to a standard algorithm  SVM implementation.  While most such quantum kernel machines are implemented using qubit based circuit architecture \cite{Havlicek2019, glick2022covariant,Schuld2021,Jerbi_2023}, our model employs continuous variable degrees of freedom of electromagnetic field modes. To date, few proposals have explored this approach to QML \cite{henderson2024quantum,Bowie_2023,Tiwari_2022_coherent,Ghobadi_PhysRevA.104.052403,Liu_2023}. We propose an implementation based on Kerr nonlinear effects in superconducting quantum circuits.

Binary classification of multi-component data elements ${\bm x}=(x_1,x_2,\dots, x_n)\in {\mathbb R}^N$ is a central task of machine learning. The overall objective is to learn a function $f:{\mathbb R}^N\rightarrow \{-1,+1\}$. However we only know this function on a finite number of points ${\bm x}_n$ the training data. We thus need to resort to an approximation scheme. 

The data is regarded as being drawn from some unknown statistical distribution.   In supervised learning we are given many examples of the data ${\bm x}_j$, together with a corresponding label $y_j=\pm 1$. We use the training set to find an unknown function such that $f({\bm x}_j)=y_j$, and use it to classify a new datum ${\bm x}$, with low probability of error.

There are a number of ways we can do this. One is based on the Support Vector Machine (SVM) algorithm (see Appendix). It first maps each datum to a vector in a higher dimensional space ${\bm x}\rightarrow {\bm \Phi}({\bm x})\in {\mathbb R}^M$, $M>N$,  and seeks a separating plane in the higher dimensional space, called the feature space, so that   $f({\bm \Phi}({\bm x}))\approx  \pm 1$.

 We  define a {\em kernel} function as an inner product in the feature vector space. 
\begin{equation}
    k({\bm x}, {\bm y}) = {\bm \Phi}({\bm x})\cdot{\bm \Phi}({\bm y})
\end{equation}
The construction indicates that the correlation matrix
\begin{equation}
    {\mathbb E}[{\bm \Phi}({\bm y})\cdot{\bm \Phi}({\bm x})]
\end{equation}
where ${\mathbb E}$ is an average over the data distribution must be positive definite if the data is distributed according to a valid probability measure. This feature ensures that $k({\bm x}, {\bm y})$  is a valid kernel. Sampling the kernel on training data ${\bm x}_j$ to construct the positive definite matrix  $K_{ij}= k({\bm x}_i, {\bm y}_j) $ can be used as input to a support vector machine algorithm to classify unlabelled test data by  assigning the label as $y={\rm sign}( d({\bm x},{\bm \alpha}_*))$ after computing the decision function
\begin{equation}
    d({\bm x},{\bm \alpha}_*)=\sum_{j=1}^n y_j \alpha_{*j}  k({\bm x}, {\bm x}_j)-b
\end{equation}
where the constants $\alpha_{*\ j},b$ are obtained through the minimisation of a function derived from the matrix $K_{ij}$.

\section{Quantum kernels.}
The passage to quantum kernels begins with parameterised quantum states, either pure or mixed \cite{schuld2021supervised}. In the unitary case we consider states of the form
\begin{equation}
    |{\bm \phi}({\bm x})\rangle = U({\bm x})|\psi_0\rangle
\end{equation}
where ${\bm x}$ is a data element and we call the state $|\psi_0\rangle$ the \textit{fiducial} state in some suitable Hilbert space. In this paper the Hilbert space is the tensor product space of $N$ bosonic modes of the electromagnetic field.

The key difference between quantum and classical states is that all quantum states are a source of randomness. In the case of pure states, there is always one physical quantity for which the measurement outcomes are completely certain and, simultaneously, there is at least one physical quantity for which the outcomes are as random as possible given other constraints (such as symmetries). The statistical distance between two parameterised pure states is simply given in terms of the inner product\cite{Wootters}
\begin{equation}
\label{q-kernel}
    k({\bm x},{\bm y}) = |\langle {\bm \phi}({\bm x})|{\bm \phi}({\bm y})\rangle\rangle|^2
\end{equation}
and we take this to define the kernel. 
More general measures are used for mixed states\cite{BRAUNSTEIN1996135}, but in this paper we will only use the unitary quantum kernel defined in Eq.(\ref{q-kernel}). Clearly the kernel thus defined is a positive definite function of the data.

If we know the quantum kernel function $k({\bm x},{\bm y})$ we can easily use an SVM algorithm that takes a custom kernel. The point of using an actual quantum device to do the training is that we never need to know the kernel function.  It is generated by making appropriate measurements on a quantum device.  The measurement data can then be passed to a support vector machine algorithm on a conventional CMOS computer with the von Neumann architecture.   This requires a shift in our perspective from algorithms to machines. There are two ways to do this. One is based on a variational circuit approach with feedback and the other is based on kernel sampling. These are the analogue of the  sequential and parallel scenarios of \cite{Goto}. In both cases we need to specify what to measure.
  


Let $\hat{\Pi}$ be an operator with zero-trace such that $\hat{\Pi}^2=1$,   for example, the parity operator for a bosonic mode $e^{i\pi a^\dagger a}$.   It follows that 
$\hat{\Pi}$ has eigenvalues equal to $\pm 1$ and must take the form 
 \begin{equation}
 \hat{\Pi}=|+\rangle\langle +|-|-\rangle\langle -| 
\end{equation}
 where $\{ |\pm\rangle\}$ are any eigenstates of $\hat{\Pi}$. The operator $\hat{\Pi}$ is self-adjoint and unitary. Define the positive operator valued measure 
\begin{equation}
    \hat{P}(y) = \frac{1}{2}(1+y\hat{\Pi})
\end{equation}
This is easily seen to be a projection operator. Thus the probability distribution for $y$ is 
\begin{equation}
    p(y) = {\rm tr}[ \hat{P}(y)\rho]
\end{equation}
where $\rho$ is an arbitrary state. The conditional state, given $y$, is 
\begin{equation}
    \rho_{|_y}= [p(y)]^{-1}\hat{P}(y)\rho\hat{P}(y)
\end{equation}
 
  The eigenvalues of $\hat{\Pi}$ are highly degenerate in a bosonic system. A unitary transformation of $\hat{\Pi}$ given by $\hat{\Pi}(\alpha)=U(\alpha) \hat{\Pi}U^\dagger(\alpha) $ is also zero-trace with $\hat{\Pi}(\alpha)^2=1$, and $\alpha\in {\mathbb C}$. If we measure $\hat{\Pi}(\alpha)$ on an encoded state $|\phi({\bm x})\rangle$ the 
probability distribution for the measurement outcomes is 
    \begin{equation}
        p(y: {\bm x},\alpha)= \langle \phi({\bm x})|\hat{P}(y,\alpha) ||\phi({\bm x})\rangle\ .
    \end{equation}
    where $\hat{P}(y,\alpha)=U(\alpha)\hat{P}(y)U^\dagger(\alpha)$.
    Defining the real function
    \begin{equation}
    d(\alpha,{\bm x})=\langle \phi({\bm x}) |\Pi(\alpha)|\phi({\bm x})\rangle
\end{equation}
we can write
  \begin{equation}
  \label{prob-outcomes}
        p(y: {\bm x},\alpha)= \frac{1}{2}(1+y d(\alpha,{\bm x})) .
    \end{equation}
    As $|d(\alpha,{\bm x})|\leq 1$ this is always positive. If $|\phi({\bm x})\rangle$ is an eigenstate of $\hat{\Pi}(\alpha)$ with eigenvalues $y$ then $p(\tilde{y}: {\bm x},\alpha)=\delta_{y,\tilde{y}}$. The objective is to find a form for $d(\alpha,{\bm x})$ that can be used as a decision function. This will require that $d(\alpha,{\bm x})$  vary non linearly between its bounds of $\pm1$ with respect to ${\bm x}$.

We will take $\alpha\in {\mathbb C}$ as drawn from a bivariate distribution ${\cal P}(\alpha)$ with zero mean, then $U(\alpha)$ is a random unitary operator,  as is $\hat{\Pi}(\alpha)$ and $\hat{P}(y,\alpha)$ with  
\begin{equation}
    \int d^2\alpha{\cal P}(\alpha) \hat{P}(y,\alpha)= \hat{P}(y)
\end{equation}
and 
\begin{equation}
    \int d^2\alpha {\cal P}(\alpha)  \hat{\Pi}(\alpha)= \hat{\Pi}
\end{equation}
from which it follows that 
\begin{equation}
\label{mixed-measurement}
     \int d^2\alpha {\cal P}(\alpha)  d(\alpha,{\bm x})= \langle \phi({\bm x})|\hat{\Pi} |\phi({\bm x})\rangle.
\end{equation}
The  move to random unitaries, or random circuits in qubit circuit based models,  has been recognised by many as a key step towards quantum machine learning\cite{random-curcuits,e25020287}.

In the case of bosonic systems we can begin with the parity operator $\hat{\Pi}=e^{i\pi a^\dagger a}$ where $[a,a^\dagger]=1$. A natural choice for  is  $d(\alpha,{\bm x})$ is a scaled version of the Wigner function,
\begin{equation}
   d(\alpha,{\bm x}) = \frac{\pi}{2} W(\alpha,{\bm x}) 
\end{equation}
where  the Wigner function evaluated at $\alpha\in {\mathbb C}$ is
\begin{equation}
 W(\alpha,{\bm x})  =\frac{2}{\pi} \langle \phi{(\bm x})| D(\alpha) \hat{\Pi}D^\dagger(\alpha)|\phi({\bm x})\rangle   
\end{equation}
and the displacement operator is defined by 
\begin{equation}
    D(\alpha) = e^{\alpha a^\dagger -\alpha a}
\end{equation}
The operator $\hat{\Pi}(\alpha)$ can be written in terms of a positive (projection) operator valued measure (POVM) $\hat{P}(y:\alpha)$. 
A measurement of $\hat{P}(y:\alpha)$ can only given the binary results $y=\{0,1\}$ regardless of the choice of $\alpha$.  However the conditional state given a measurement of $\hat{P}(y:\alpha)$  is given by
\begin{equation}
    |\phi({\bm x}|y:\alpha)\rangle=[p(y:{\bm x}, \alpha)]^{-1/2}\hat{P}(y:\alpha)|\phi({\bm x})\rangle 
\end{equation}
with 
   \begin{equation}
p(y: {\bm x}, \alpha)=\frac{1}{2}(1+ y\frac{\pi}{2} W(\alpha,{\bm x})) 
   \end{equation}
   Using Eq. (\ref{mixed-measurement}) we see that if we average the decision function over $\alpha$ it becomes the scaled Wigner function of the encoded state  at the origin. 
   \begin{equation}
       \int d^2\alpha  {\cal P}(\alpha) d(\alpha,{\bm x})= \frac{\pi}{2} W(0,{\bm x})
   \end{equation}
   This shows that we are sampling the Wigner function of the encoded state near the origin. 
   
   The appearance of the Wigner function here is natural due to its role in Wigner state tomography\cite{Royer}. The connection between the Wigner function and the decision function   $d(\alpha,{\bm x}) $ immediately implies that we require non-classical bosonic states for quantum kernel learning. Recall, for our binary classification task, the decision function must return both positive and negative values to correctly classify new data points. Given Wigner functions for all classical states are positive, learning is the manner we describe here cannot be achieved without a quantum encoding. Such Wigner negative states can be generated by Kerr non linearities—in the following section we introduce such an encoding.

\section{The Kerr quantum kernel.}
What is the unitary, $U(\vec{x})$, and what is the fiducial state, $|\phi_0\rangle$, best suited to the learning task at hand? The answer to the first question comes down to defining generators of the unitary operator. We are interested in the quantum states of $P$ bosonic modes associated with annihilation and creation  operators $a_p, a_p^\dagger$ with $p=1,2,\ldots P$.
Lloyd and Braunstein \cite{PhysRevLett.82.1784} have shown that a universal simulation of a  single bosonic mode is generated by the algebra 
\begin{equation}
    \{ a, a^\dagger , a^2, a^{\dagger\ 2}, (a^\dagger a)^2\}
\end{equation}
The first two operators generate displacements (Heisenberg-Weyl group) the next two quadratic operators generate the squeezing transformations (SU(1,1)). The quartic operator is called the Kerr nonlinearity.  In this paper we select  the fiducial state from the class of displaced squeezed states, and select unitaries generated by combinations of the multi-mode Kerr generators $\{(a^\dagger_ja_j)^2, 2a_j^\dagger a_j a_k^\dagger a_k\}$ for $j,k=1,2,\ldots P$ and $j\neq k$.

The quantum kernel state is defined by
\begin{equation}
|{\bm \phi}({\bm x})\rangle = U({\bm x}) |{\bm \beta}, {\bm r}\rangle
\end{equation}
where $U({\bm x})$ is a function of Kerr non linear generators and $|{\bm \beta}, {\bm r}\rangle =|\beta_1,r_1\rangle\otimes|\beta_2,r_2\rangle\otimes\ldots |\beta_P,r_P\rangle$ 
is a tensor product of independent displaced squeezed states defined in the number basis as\cite{WM} 
\begin{equation}
 \langle n|\beta,r\rangle \equiv f_n(\beta, r)=\langle n|\beta,r\rangle= (n!\cosh r)^{-1/2}\lambda^{n/2}e^{-|\beta|^2/2-\lambda \alpha^2}H_n(z)
  \end{equation}
  where
  \begin{equation}
  z=\sqrt{\lambda}\beta+\frac{\beta^*}{2\sqrt{\lambda}}
  \end{equation}
and $\lambda=\frac{1}{2}\tanh r$\ ,
with $|0\rangle$ the $P$-mode vacuum state.

As in the introduction we define a vector of data parameters ${\bm x}=(x_1,x_2,\ldots x_n)\in {\mathbb S}^P$, where ${\mathbb S}$ is the unit interval,  and a vector-valued map:  ${\bm \phi}: {\mathbb R}^N\rightarrow {\mathbb R}^K\ \mbox{with }K\geq N$ such that each component is a function of ${\bm \phi}({\bm x}) \in{\mathbb R}^K$. The $K=P(P+1)/2$ components of ${\bf \phi}$ are
\begin{eqnarray}
    \phi_k & = & x_k^2\ \ \  k=1,2,\ldots P\\
    \phi_{P+k} & = & (1-x_k)(1-x_{k+1})\ \ \  k=1,\ldots P(P-1)/2
\end{eqnarray}
The  $K=P(P+1)/2$ generators  are defined as 
\begin{eqnarray}
    \hat{g}_k  & = &  \hat{n}_k^2, \ \ \  k=1,2,\ldots P\\
    \hat{g}_{P+k} & = & 2\hat{n}_k\hat{n}_{k+1}\ \ \  k=1,\ldots P(P-1)/2
\end{eqnarray}
and the unitary is defined as
\begin{equation}
U({\bm \phi}({\bm x}) =e^{-i\pi {\bm \phi}({\bm x})\cdot \hat{{\bm g}}}
\end{equation}
For example, in the case of two modes, 
\begin{equation}
    {\bm \phi}({\bm x})\cdot \hat{{\bm g}}= x_1^2 (a_1^\dagger a_1)^2+x_2^2 (a_2^\dagger a_2)^2+ 2(1-x_1)(1-x_2) a_1^\dagger a_1 a_2^\dagger a_2\ .
\end{equation}

The fiducial states are set with the same values $\alpha_k=\alpha_0,r_k=r_0$.  The state encoding is then 
\begin{equation}
|{\bm \phi}({\bm x})\rangle = e^{-i\pi {\bm \phi}\cdot {\bm \hat{\bm g}}}|\alpha_0,r_0\rangle
\end{equation}

In the number state basis we can write
\begin{equation}
|{\bm \phi}({\bm x})\rangle= \sum_{{\bm n}}^\infty f_{\bm n}(\alpha_0, r_0) e^{-i\pi {\bm 
\phi}({\bm x}).{\bm g}({\bm n})}|{\bm n}\rangle
\end{equation}
where ${\bm n}^T =(n_1,n_2,\ldots, n_P)$ and the sum means   
\begin{equation}
    \sum_{{\bm n}}^\infty= \sum_{n_1=0}^\infty\sum_{n_2=0}^\infty \ldots \sum_{n_P=0}^\infty
\end{equation}
while $|{\bm n}\rangle=\prod_j^{\otimes}|n_j\rangle$ and $ f_{\bm n}(\alpha_0, r_0)=\prod_{ n_j} f_{n_j}(\alpha_0,r_0)$.  The functions ${\bm g}({\bm n})$ are the same as the operator valued functions with operators replaced by integers.  It is now easy to see that the states $|{\bm \phi}({\bm x})\rangle$ are not orthogonal and in fact over-complete 
\begin{equation}
    (2\pi)^{-P}\int d{\bm x} \ |{\bm \phi}({\bm x})\rangle\langle {\bm \phi}({\bm x})|={\mathbb I}
    \label{overcomplete}
\end{equation}

We define the kernel function as 
\begin{equation}
\label{kerr-kernel}
k({\bm y},{\bm x}) = |\langle {\bm \phi}({\bm y})|{\bm \phi}({\bm x})\rangle|^2
\end{equation}\
It follows from Eq. (\ref{overcomplete})  that 
\begin{equation}
  \int d{\bf x}\  k({\bm y},{\bm x}) =1
\end{equation}

\subsection{Example: one mode case. }
Let us now consider a single real parameter $x$ confined to the unit interval, $0 \leq x\leq 1$.  We define a single parameter quantum encoding using 
\begin{equation}
\label{single-mode-encode}
    |\phi(x)\rangle =  e^{-i\pi x (a^\dagger a)^2}|\alpha_0\rangle 
\end{equation}
where we have taken the fiducial state to be a fixed coherent state $|\alpha_0\rangle$. The corresponding kernel is 
\begin{equation}
\label{one-mode-kernel}
k(y,x)= |\langle \alpha_0 |e^{-i\pi (x-y) (a^\dagger a)^2}|\alpha_0\rangle |^2
\end{equation}
Using the number state representation of a coherent state
\begin{equation}
  k(y,x)= e^{-2|\alpha_0|^2}\left | \sum_{n=0}^\infty \frac{|\alpha_0|^2}{n!} e^{-i\pi(x-y)n^2}\right |^2
\end{equation}
This function is related to the trace of the unitary operator generated by the Kerr interaction and the curlicule functions discussed by Berry and Goldberg\cite{Berry}. 

The decoding is done by measurement of the displaced parity operator defined by 
\begin{equation}
\label{dis-parity}
    \hat{\Pi}(\mu) =  D (\mu)e^{-i\pi a^\dagger a}D^\dagger(\mu)
\end{equation}
where $D(\mu)$ is a displacement operator in a complex random variable $\alpha$, defined by 
\begin{equation}
    D(\mu) = e^{\mu a^\dagger -\mu^* a}
\end{equation}
The average value of the displaced parity operator on the encoded state is function
\begin{equation}
   d(\mu,x) = \langle \phi(x)| D(\mu) \hat{\Pi}D^\dagger(\mu)|\phi(x)\rangle   
\end{equation}
This is  $\pi/2$ times the Wigner function for the encoded state, evaluated at $\alpha$. 

The function $d(\mu,x)$ is called the decision function. As above, this expectation value can only become negative for non-classical states, in particular, for the encoded states in Eq.(\ref{single-mode-encode}). It is straightforward to measure the displaced parity operator, Eq. (\ref{dis-parity}), on an arbitrary state as we discuss in section (\ref{exp}).

We can evaluate the average for an arbitrary state 
\begin{equation}
    \langle \psi|\hat{\Pi}(\mu)|\psi\rangle =  \sum_{n=0}^\infty (-1)^n|\langle n|D(\mu)|\psi\rangle|^2
\end{equation}
where 
\begin{equation}
    \langle \psi|D(\mu)|n\rangle=\sum_{m=0}^\infty \psi_m D_{mn}(\mu)
\end{equation}
with $\psi_m=\langle \psi|m\rangle$ and 
\begin{equation}
\label{displacement-number-basis}
    D_{mn}(\mu)= \sqrt{\frac{n!}{m!}}e^{-|\mu|^2/2}z^{m-n} L_n^{m-n}(|\mu|^2)\ \ \ \ \ m\geq n
\end{equation}
where $L_p^q$ are the associated Laguerre polynomials.

Inserting the resolution of identity,  $I=\sum_k|k\rangle\langle k|$ we can write this 
\begin{equation}
\label{alternating}
    d(\mu,x)=\sum_{k=0}^\infty (-1)^k  g(k,\mu,x)
\end{equation}
where 
\begin{equation}
\label{displaced-number}
 g(k,\mu,x)= \left |\langle \phi(x)|D(\mu)|k\rangle\right |^2
 \end{equation}
 This is simply the probability of making a measurement of photon number on the displaced state $D(-\mu)|\phi(x)\rangle$ and finding the result $k$.  It follows that
 \begin{equation}
     0 <g(k,\alpha,x)< 1
 \end{equation}

 Writing the state in the number basis
 \begin{equation}
     |\phi(x)\rangle =e^{-|\alpha_0|^2/2}\sum_{n=0}^\infty \frac{\alpha_0^{n}}{\sqrt{n!}} e^{-i\pi x n^2}|n\rangle 
 \end{equation}
 we get 
 \begin{equation}
     g(k,\alpha,x)=e^{-|\alpha_0|^2}\left |\sum_{n=0}^\infty \frac{\alpha_0^{n}}{\sqrt{n!}} e^{i\pi x n^2} d_{nk}(\alpha)\right |^2
 \end{equation}
 where the matrix elements of the displacement operator in the number basis are
 \begin{equation}
 \label{dispalce-matrix}
   d_{nm}(\alpha)  = \langle n|D(\alpha)|m\rangle=\sqrt{\frac{m!}{n!}}e^{-|\alpha|^2/2}\alpha^{n-m} L_m^{n-m}(|\alpha|^2) \ \ \ \ n\geq m\ \ ,
 \end{equation}

\begin{figure}[htbp!]
    \centering
    \includegraphics[scale=0.5]{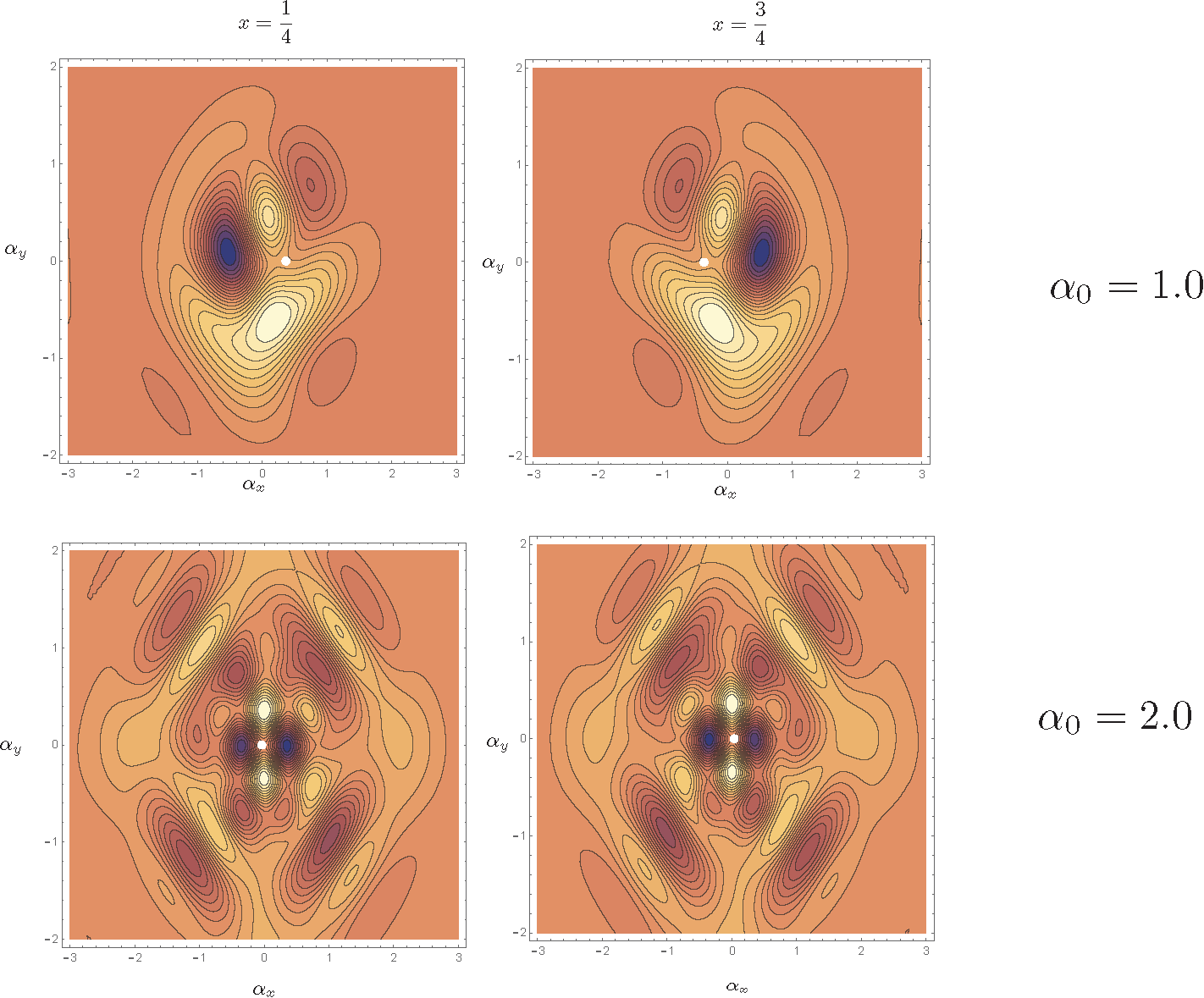}
    \caption{The Wigner function contours for a single mode cat states with $\alpha_0=1.0,2.0$, $x=1/4, x=3/4$. Blue indicates regions of negativity. The coordinates are defined by the displacement $\alpha=\alpha_x+i\alpha_y$. The white dot indicates the bias point defined in Eq.(\ref{bias-func}). }
    \label{new-WignerKC}
\end{figure}

\begin{figure}[htbp!]
    \centering
    \includegraphics[scale=0.5]{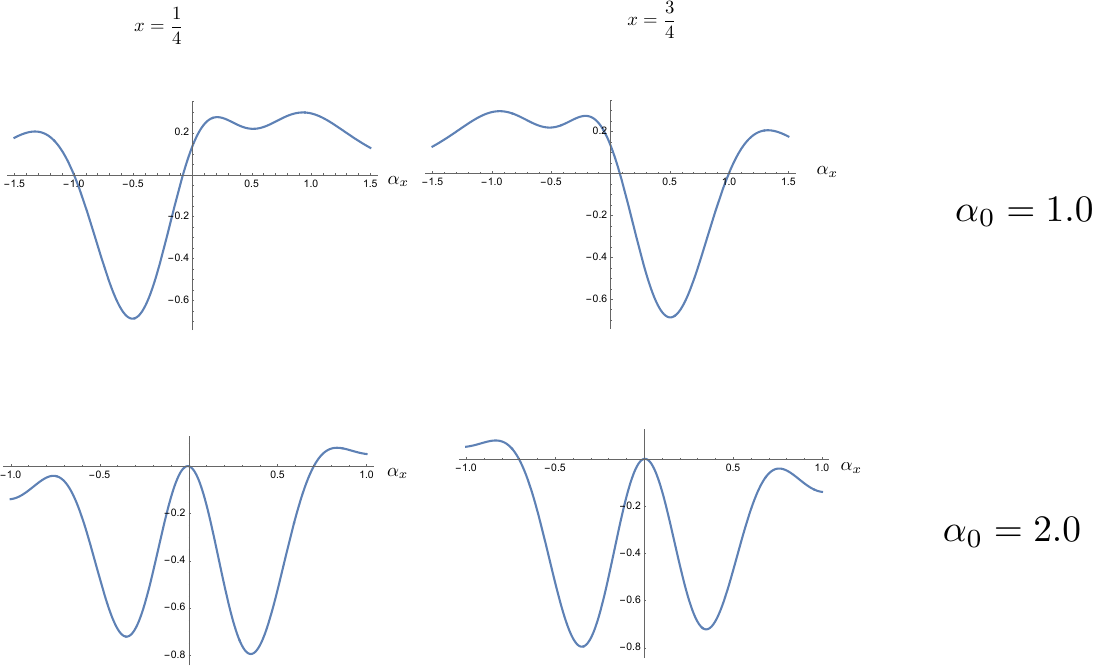}
    \caption{The Wigner function cross section for $\alpha_y=0$ for a single mode cat states with $\alpha_0=1.0,2.0$, $x=1/4, x=3/4$.  }
    \label{wigner-minima}
\end{figure}

 We now turn to a classification problem. We construct a specific example to illustrate the key features of this learning scheme. Assume that the data points are rational numbers, $x=p/q$ and define the labelling function
\begin{equation}
 f(x) = {\rm sign} [\cos(x\pi)]
\end{equation}
As an example we will use the two cases $x=1/4$ and $x=3/4$. The decision boundary is trivial and occurs at $x=1/2$ on the unit interval.   In this case the encoded states are simply the two Kerr cat states 
\begin{eqnarray}
    |\phi(1/4)\rangle & = & \frac{1}{2}\left (|i\alpha_0\rangle +|-i\alpha_0 \rangle\right ) +\frac{e^{-i\pi/4}}{2}\left (|\alpha_0\rangle-|-\alpha_0\rangle\right )\\
    |\phi(3/4)\rangle & = & \frac{1}{2}\left (|i\alpha_0\rangle +|-i\alpha_0 \rangle\right ) +\frac{e^{i\pi/4}}{2}\left (|\alpha_0\rangle-|-\alpha_0\rangle\right )  
\end{eqnarray}
These states are  not orthogonal
\begin{equation}
    \langle \phi(3/4)|\phi(1/4)\rangle =(1+e^{-4|\alpha_0|^2})/4
\end{equation}
These are plotted in the top row of Fig. (\ref{new-WignerKC}). 
In Fig. (\ref{wigner-minima}) we plot the Wigner function along the real axis  for the case shown in Fig.(\ref{new-WignerKC}). The two cases are distinguished by the location of the minima along this axis.

\begin{figure}
    \centering
    \includegraphics[scale=0.5]{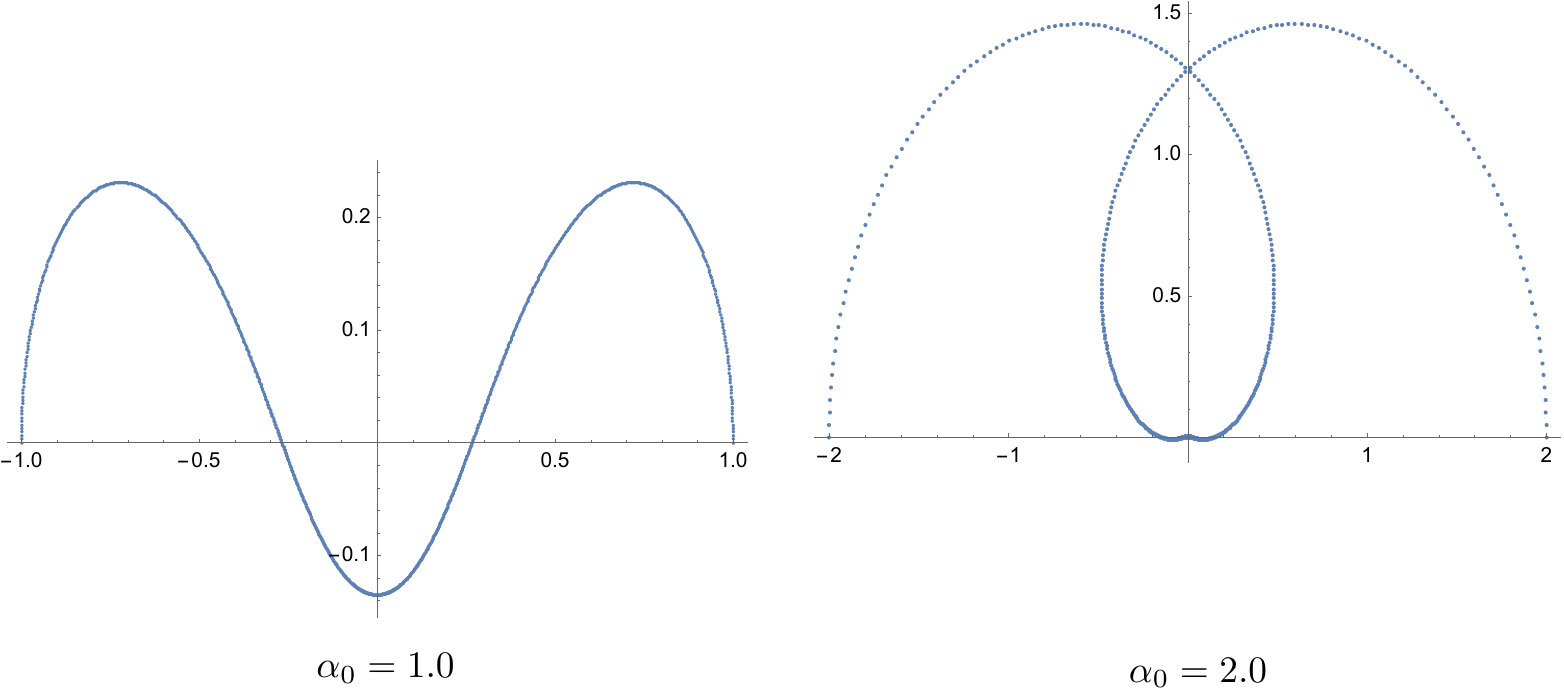}
    \caption{The mean value of the amplitude as a parametric function of $x$ on the unit interval for two different values of $\alpha_0$. }
    \label{bias} 
\end{figure}

 An important role is played by the average amplitude
\begin{equation}
\label{bias-func}
   b({\bm x})= \langle \phi({\bm x})|a|\phi({\bm x})\rangle=e^{-i\pi x}e^{-|\alpha_0|^2(1-e^{-2\pi i x})}
\end{equation}
In Fig. (\ref{bias}) we plot this in the complex plane as a parametric function of $x$ on the unit interval for two values $\alpha_0=1.0,2.0$. The collapse and revival familiar from Kerr nolinearities is evident\cite{HolMil}. 
In the two cases in Fig. (\ref{new-WignerKC}) this is given by
\begin{equation}
   b(1/4) = \alpha_0 e^{-i\pi/4} e^{-|\alpha_0|^2(1+i)} ;\ \ \ \ \ \  b(3/4) = \alpha_0 e^{i\pi/4} e^{-|\alpha_0|^2(1-i)}
\end{equation}
These are complex conjugate pairs and thus are transformed into each other by reflection around the $\alpha_x$ axis in the complex plane. They are shown as a white cross in Fig. (\ref{new-WignerKC}). 


We will use a supervised learning scheme based on labelled data sets. This means that we have a large training set of labelled data points $(x,f(x))$. In accordance with the usual ML nomencalture, we would set $f(x)=y$ and refer to $y$ as the true training label for $x$. We will assume that both labels occur with equal probability in the training data.  As stated in the introduction, in the theory of quantum learning there are two scenarios we can use that we refer to as the parallel and sequential\cite{Goto}. These two protocols are also used in \cite{Havlicek2019} where they are called the quantum kernel estimator and the quantum variational classier respectively. However that approach encodes quhits into a superconducting circuit whereas here we propose using CV analogue quantum processing. 

\subsubsection{Sequential protocol.}
In an experimental setting we can feed the labelled training data into the Kerr nonlinearity using a time series of labelled data points. The input signal to the Kerr non linearity is thus a square-wave signal with values $x=(1/4,3/4)$ and the corresponding labels follow with values $(+,-)$ as shown in the figure. This is called the 'sequential learning protocol'. This is the CV equivalent of a variational quantum circuit (VQC) for qubit encoding. 
\begin{enumerate}
\item[Step 1] Using a Kerr Hamiltonian, make the state $|\phi(x)\rangle$ with true label $y$.
$$
|\phi(x)\rangle= U(x)|\alpha_0\rangle
$$
\item[Step 2] Displace the state by a small random complex number $-\mu\in {\mathbb C}$
$$
|\phi(x)\rangle\rightarrow D^\dagger(\mu)|\phi(x)\rangle
$$
\item[Step 3] Measure the parity operator $\Pi$ to get the result $\tilde{y}=\pm 1$
\item[Step 4] Compute the error, which will always be 0 or 1 
\begin{equation}
    \epsilon =\frac{1}{4}(y-\tilde{y})^2
\end{equation}
\end{enumerate}
We repeat this $K$ times, with the same input state and same label,  and compute the average error
\begin{equation}
    \bar{\epsilon}=\frac{1}{K}\sum_{j=1}^K \epsilon_j=\frac{1}{2}(1-y\bar{y})
\end{equation}
where we have used the fact that in each trial $\tilde{y}_j=\pm 1$ and $y$ is fixed at $\pm 1$. In the case of many trials $K>>1$ this should converge to 
\begin{equation}
\label{average-error}
    \bar{\epsilon}=\frac{1}{2}(1-yd(\mu,x))
\end{equation}

We refer to this $K$ trials as an {\em epoch}. It enables us to sample the average error through empirical means.


We now feedback to change the random displacement for the next epoch to $\mu_{k+1}=\mu_k+\delta_k$ in such a way as to {\em decrease} the average error in each epoch. Define $\mu=u+iv$ then write the update rule as
\begin{eqnarray}
     u_{k+1}  & = & u_k + \Delta u_k\\
      v_{k+1}  & = & v_k + \Delta v_k
\end{eqnarray}
  
The  first-order change in the average error is determined by
\begin{equation}
    \Delta\bar{\epsilon} = \partial_{u} \bar{\epsilon}\  \Delta u+\partial_{v} \bar{\epsilon}\  \Delta v 
\end{equation} 
with 
\begin{equation}
   \partial_u=\frac{\partial }{\partial u}\ ,\ \ \   \partial_{v}=\frac{\partial}{\partial v}\  
\end{equation}
If we now choose the update rule as 
\begin{equation}
\label{update}
    \Delta u_k=-\partial_u \bar{\epsilon}\ ,\ \ \ \ \Delta v_k=-\partial_v \bar{\epsilon}
\end{equation}
we see that the change in the error is 
\begin{equation}
    \Delta\bar{\epsilon}= -\left [(\partial_u \bar{\epsilon})^2+(\partial_{v} \bar{\epsilon})^2\right ]
\end{equation}
This says that the error decreases from one epoch to the next and that to make the change in the error as large as possible we need to maximise 
\begin{eqnarray}
    G(\mu,x) & = & \left [(\partial_u \bar{\epsilon})^2+(\partial_{v} \bar{\epsilon})^2\right ]\\
           & = & \left (\frac{\partial\bar{\epsilon}}{\partial \mu}\right )\left (\frac{\partial\bar{\epsilon}}{\partial \mu^*}\right )\\
            & = &\left |\frac{\partial\bar{\epsilon}}{\partial \mu}\right |^2
\end{eqnarray}
In other words we need to find the steepest gradient descent of the error in the complex plane. 

Using Eq. (\ref{average-error}) we see that the update rule is 
\begin{equation}
    \Delta\mu_k=-\frac{y}{2} (\partial_u +i \partial_v)d(\mu,x)= -y\partial_{\mu^*} d(\mu,x)\ , 
\end{equation}
and 
\begin{equation}
    \Delta\bar{\epsilon}= -\frac{1}{4}\left |\frac{\partial d(\mu,x)}{\partial\mu}\right |^2
\end{equation}
In the experiment we use the epoch sampled (empirical) value for $d(\mu,x)$ which fluctuates from one epoch to the next. Thus the displacements become a stochastic process in the complex plane, where the stochasticity is driven by the probabilistic nature of quantum measurement. As $d(\mu,x)$ is a rescaled Wigner function of the encoded state as a function of $\mu$ we see that we are using a gradient descent algorithm on the Wigner function for $|\psi(x)\rangle$.

Using Eq. (\ref{prob-outcomes}) we see that the update rule is 
\begin{equation}
      \Delta \mu_k=-2y\frac{\partial p(\mu_k,x)}{\partial \mu^*_k}
\end{equation}  
and 
\begin{equation}
     \Delta\bar{\epsilon}=- \left |\frac{\partial p(\mu,x)}{\partial \mu^*}\right |^2
\end{equation}
where 
\begin{equation}
    p(\mu,x)=   \langle \phi(x)| D(\mu) \hat{P}D^\dagger(\mu)|\phi(x)\rangle 
\end{equation}
and 
\begin{equation}
    \frac{\partial p(\mu,x)}{\partial \mu^*}=\mu(1-p(\mu,x))-b(x)
\end{equation}
where $b(x) = \langle \phi(x)|a|\phi(x)\rangle$ is the mean amplitude of the encoded state. When  $x=1/2$ or $x=3/2$  the mean amplitude is zero (symmetric cat) and we can neglect this term.   Thus we have 
\begin{eqnarray}
     \Delta \mu_k & = & -2y\mu_k(1-p(\mu_k,x))=-y\mu_k(1-d(\mu_k,x))\\
     \Delta\bar{\epsilon}_k& = & -|\mu|^2(1-p(\mu_k,x))^2= -|\mu_k|^2(1-d(\mu_k,x))^2/4
\end{eqnarray}
where we have used $p(\mu,x) = (1+d(\mu,x))/2$ and the decision function $d(\mu,x)$  is given in Eq. (\ref{alternating})

In the special case of $|\phi(1/2)\rangle$ and $|\phi(3/2)\rangle$ we find that to lowest order in $\mu_k$ ,
\begin{eqnarray}
\label{updates}
     \Delta \mu_k & = & \mp 2y\alpha_0\mu_k \Im(\mu_k)\\
     \Delta\bar{\epsilon}_k& = & -\alpha^2_0|\mu_k|^2\Im(\mu_k)^2
\end{eqnarray}
We see that as optimisation proceeds through feedback $p\rightarrow 1$ and  the displacements of the parameters and the error stops changing. This is what one would expect for this kind of learning\cite{CP-LM-review}.

However it is already clear that the single mode Kerr is distinctly different from the corresponding classical learning problem precisely because it generates states for which the Wigner function is negative somewhere in the complex plane. Instead of a real weight parameter we have a complex weight parameter.

\subsubsection{Parallel protocol (quantum kernel method).}
In this method we sample the kernel $k(x,x')$ using measurements and use the corresponding values in a support vector machine algorithm to find the support vectors.  Let  $x_1, x_2,\dots x_K$  be a discrete sampling of the data with the corresponding true labels $y_1,y_2,\dots y_K$.  Likewise for $x'_k,y'_k$.  The kernel, often referred to as a fidelity kernel, is given by 
\begin{equation}
    k(x_j,x_k) =|\langle 0|D^\dagger(\alpha_0) |U^\dagger(x_j) U(x_k)D(\alpha_0)|0\rangle|^2.
\end{equation}
we see that the experimental protocol will apply the unitary operator $D^\dagger(\alpha_0) |U^\dagger(x_j) U(x'_k)D(\alpha_0)$ to the vacuum state and then measure the photon number. The value of the kernel is then simply the probability that the photon number is zero. In an experiment, we need to sample this probability over many trials and estimate the matrix elements $K_{jk}=k(x_j,x_k)$.

To simulate an experiment, we use the single-mode analytic expression, Eq. (\ref{one-mode-kernel}) to compute $ k(x_j,x_k)$ for $x_j,x_k$ drawn randomly from the unit interval. This is a probability and may be sampled as follows. For each pair $(x_j,x_k)$ choose a random number $0< r < 1$. If $r>  k(x_j,x_k)$ record $0$, else record $ 1 $. Repeat $M$ times to find $Pr(0)=m(0)/M$ where $m(0)$ is the number of trials that given $0$. This estimates the value of the kernel matrix element,  $K_{jk}$ with an error that scales as $M^{-1/2}$.  Once the matrix $K$ has been sampled we can implement  the SVM algorithm. We will not do this here and move on to the two mode case which explicitly uses entanglement.





\subsection{Example: Two mode case.}
The essential change in moving to the two mode case is the introduction of entanglement between the two modes.
Two-mode entangled Kerr-cats are generated by a general two-mode Kerr transformation of a product state of two identical coherent states $|\alpha_0,\alpha_0\rangle=|\alpha_0\rangle \otimes|\alpha_0\rangle$  as
\begin{equation}
    |\phi({\bm x})\rangle =e^{-i\pi (x_1 (a^\dagger a)^2+x_2 (b^\dagger b)^2+2x_{3} a^\dagger a b^\dagger b)}|\alpha_0,\alpha_0\rangle
\end{equation}
where ${\bm x}=(x_1,x_2,x_3)$ and with $x_j\in [0,1]$.
The number state basis representation is 
\begin{equation}
|\phi({\bm x})\rangle =e^{-|\alpha_0|^2}\sum_{n,m=0}^\infty \frac{\alpha_0^n\alpha_0^m}{\sqrt{n!m!}}\exp\left [-i\pi n( nx_1+mx_3)-i\pi m( mx_2+n x_3)\right ]|n, m\rangle
\end{equation}
The reduced state of either mode is 
\begin{equation}
  \rho= \sum_{n=0}^\infty \frac{|\alpha_0|^{2n}}{n!}e^{-|\alpha_0|^2} |\alpha_0 e^{-2\pi i x_3 n}\rangle\langle  \alpha_0 e^{-2\pi i x_3 n}|
\end{equation}
The two-mode Kerr state is separable if $x_3=0$ and $x_3=1$. When $x_3=1/r$ with $r\in{\mathbb N}$ the reduced state is entangled.   We will assume that $x_3=x_1x_2$. The state is entangled on the hyperbola $x_1x_2=1/r$. The Wigner function of the reduced state is positive and has $r$ peaks. When ${\bm x}=(1,1/2)$ we get the the two-mode cat state,
\begin{equation}
    |\phi(1,1/2)\rangle\sim e^{i\pi/4}|-\alpha_0\rangle \big [|\alpha_0\rangle+|-\alpha_0\rangle\big]+ e^{-i\pi/4}|\alpha_0\rangle\big [|\alpha_0\rangle-|-\alpha_0\rangle\big ]
\end{equation}
(up to normalisation).
On the other hand, when  ${\bm x}=(1/2,1)$ we get 
\begin{equation}
    |\phi(1/2,1)\rangle\sim e^{i\pi/4} \big [|\alpha_0\rangle+|-\alpha_0\rangle\big] |-\alpha_0\rangle+ e^{-i\pi/4}\big [|\alpha_0\rangle-|-\alpha_0\rangle\big ]|\alpha_0\rangle
\end{equation}
Tracing out either mode gives an equal mixture of two coherent states $|\pm \alpha_0\rangle$. We can distinguish these states by a variational displacement and measurement feedback as for the single mode case.    

The Wigner function for the two-mode state $|\phi({\bm x})\rangle $ is 
\begin{equation}
    W(\alpha,\beta) =\frac{4}{\pi^2}\sum_{n,m=0}^\infty (-1)^{(n+m)}p_{n,m}(\alpha,\beta)
\end{equation}
where 
\begin{equation}
  p_{n,m}(\alpha,\beta)=\left |\langle \phi({\bm x})|D(\alpha)\otimes D(\beta)|n,m\rangle  \right |^2
\end{equation}
Using the number representation for $|\phi({\bm x})\rangle$
\begin{align}
\ket{\Phi({\bm x})} &= e^{-|\alpha|^2}\sum^{\infty}_{n,m=0} \frac{\alpha^n \alpha^m}{\sqrt{n!m!}} e^{\blue{-}i\pi[n^2x_1 + m^2x_2 + 2nmx_1x_2]}\ket{n,m} \label{eq:coherentencoding}
\end{align}
we find that
\begin{equation}
 \langle \phi({\bm x})|D(\alpha)\otimes D(\beta)|n,m\rangle=\sum_{j,k=0}^\infty \phi^*_{jk}({\bm x})d_{jn}(\alpha)d_{km}(\beta)   
\end{equation}
with $d_{kn}(z)$ given by Eq. (\ref{dispalce-matrix}).  The kernel is 
\begin{equation}\label{eq:coherentkernel}
k({\bm x},{\bm y}) = e^{-2|\alpha|^2}\left|\sum^{\infty}_{n,m,=0}\frac{|\alpha|^{2(n+m)}}{n!m!} e^{i\pi\left[n^2(y_1-x_1)+m^2(y_2-x_2)+2nm\left(y_1y_2-x_1x_2\right)\right]}\right|^2.
\end{equation}
 

More generally we can use a displaced squeezed state as the fiducial state. This addds an additional parameter to the encoding .   A squeezed version of the above encoding state:
\begin{equation}
\ket{\Phi_S({\bm x})} = \hat{U}({\bm x})\left(\ket{\alpha,r}_a \otimes \ket{\alpha, r}_b \right),
\end{equation}
where $\hat{U}({\bm x})$ has the same form as before, and $r$ is the squeezing parameter.

Such a squeezed state in the number state basis takes the form:
\begin{equation}
\ket{\alpha,r} = \sum_{n=0}^{\infty}f_n(\alpha_0,r_0)e^{-i\pi\theta(x)}\ket{n},
\end{equation}

\begin{equation}
f_n(\alpha_0,r_0)=(n! \cosh r)^{-1/2}\left(\frac{\sinh r}{2 \cosh r}\right)^{n/2}e^{-\frac{|\alpha|^2}{2}-\frac{\sinh r}{2 \cosh r}\alpha^2} H_n(z),
\end{equation}
for 

\begin{equation}
z=\sqrt{\frac{\sinh r}{2 \cosh r}}\alpha + \frac{\alpha^*}{2\sqrt{\frac{\sinh r}{2 \cosh r}}}
\end{equation}

and $H_n$ are Hermite polynomials.
\medskip

Our full reference squeezed state, over two modes, is then:
\begin{equation}\label{eq:squeezedstate}
\ket{\Phi_S({\bm x})} = \sum_{n,m}^{\infty}f_m(\alpha_0,r_0)f_n(\alpha_0,r_0)e^{-i\pi\left[n^2x_1 + m^2x_2 + 2nmx_1x_2\right] } \ket{n,m}.
\end{equation}

The corresponding kernel, the square of the absolute value of the inner product $|\braket{\Phi_S({\bm y })|\Phi_S({\bm x})}|^2$, is:
\begin{equation}\label{eq:squeezedkernel}
k({\bm x},{\bm y}) = \left|\sum^{\infty}_{n,m,=0}f_m(\alpha_0,r_0)f_n(\alpha_0,r_0) e^{i\pi\left[n^2(y_1-x_1)+m^2(y_2-x_2)+2nm\left(y_1y_2-x_1x_2\right)\right]}\right|^2.
\end{equation}

\section{Quantum Kernel Machine Learning with two modes.}
\subsection{Generating the data}
Data is generated as random points on a unit square, resulting in an array of vectors $\tilde{x} = (x_1,x_2) \in [0,1]^2$.
The same 100,000 point data set was used throughout the following analysis after this initial random generation.

\subsection{Encoding and labelling the data}
To encode our data using the states discussed above, the randomly generated  $x_1$ and $x_2$ values are passed into Equations~\eqref{eq:coherentencoding} or~\eqref{eq:squeezedstate}, and a displaced parity operator is applied. The expectation value of this parity operator is then used to label the data. The eigenvlaues of teh two-mode parity operator,
which has eigenvalues $\lambda = \pm 1$, will  label the data.

We take as our displacement operator
\begin{equation}
D(\mu, \nu) = e^{\mu a^{\dagger}-\mu^* a} e^{\nu b^{\dagger}-\nu^* b},
\end{equation}
where $\mu, \nu \in \mathbb{C}$ are independent Gaussian distributed random variables, and $\mu = \mu_x + i\mu_y$ and $\nu = \nu_x +i\nu_y$, one for each of the modes. The quantities $\mu_x, \mu_y, \nu_x, \nu_y$ are independent Gaussian random variables of mean zero and unit variance.
It is important that they are restricted to the unit disk in the complex plane, to minimise the distance between encoded data points.

The labelling function/expectation value goes as:
\begin{align}
\mathcal{L}(\tilde{x}) &= \bra{\Phi(\tilde{x})}U^{\dagger}(\tilde{x})D^{\dagger}(\mu, \nu) e^{i\pi(a^{\dagger}a+b^{\dagger}b)}D(\mu, \nu)U(\tilde{x})\ket{\Phi(\tilde{x})}\\
&= \bra{\Phi(\tilde{x})}D^{\dagger}(\mu, \nu) \hat{\Pi} D(\mu, \nu)\ket{\Phi(\tilde{x})}\\
&= \bra{\Phi(\tilde{x})}\hat{\Pi}_{\mu,\nu}\ket{\Phi(\tilde{x})} \label{eq:labelling}
\end{align}

Deriving the explicit form of the labelling function:
\begin{align}
\bra{\Phi(\tilde{x})}\hat{\Pi}_{\mu,\nu}\ket{\Phi(\tilde{x})} &= \bra{\Phi(\tilde{x})}D^{\dagger}(\mu, \nu) \hat{\Pi} ~D(\mu, \nu)\ket{\Phi(\tilde{x})} \nonumber \\
&= \bra{\Phi(\tilde{x})}D^{\dagger}(\mu, \nu) ~\hat{\Pi} \sum^{\infty}_{n,m=0}\ket{n,m}\bra{n,m} D(\mu, \nu)\ket{\Phi(\tilde{x})} \nonumber \\
&= \bra{\Phi(\tilde{x})}D^{\dagger}(\mu, \nu) \sum^{\infty}_{n,m=0}(-1)^{(n+m)}\ket{n,m}\bra{n,m} D(\mu, \nu) \ket{\Phi(\tilde{x})} \nonumber \\
&= \sum^{\infty}_{n,m=0}(-1)^{(n+m)}\left|\bra{n,m} D(\mu, \nu) \ket{\Phi(\tilde{x})}\right|^2
\end{align}

Taking just the expression inside the absolute value brackets:
\begin{align}
\bra{n,m} D(\mu, \nu) \ket{\Phi(\tilde{x})} &= \bra{n,m} \sum_{jk} \ket{j,k}\bra{j,k} D(\mu,\nu) \ket{\Phi(\tilde{x})} \label{eq:matrixelements}
\end{align}


The two terms that make up $D(\mu,\nu)$ act individually on their respective modes. We hence write $\bra{n} D(\mu) \ket{j}=d_{nj}(\mu)$ and $\bra{m} D(\nu) \ket{k}=d_{mk}(\nu)$, and can further simplify Eq.~\eqref{eq:matrixelements} to
\begin{align}
\bra{\Phi(\tilde{x})} D(\mu, \nu) \ket{n,m} &= \sum_{jk}\Phi_{jk}(\tilde{x})d_{nj}(\mu)d_{mk}(\nu),
\end{align}
where $\Phi_{jk}(\tilde{x})$ appears because $\braket{j,k|\Phi(\tilde{x})} = \Phi_{jk}(\tilde{x})$.

So $\ket{\Phi(\tilde{x})}$ in Equation~\eqref{eq:labelling} can be substituted for either the Eq.~\eqref{eq:coherentencoding} coherent or the Eq.~\eqref{eq:squeezedstate} squeezed state depending on the encoding desired.

In practise, it is not tractable to have infinite sums, so we must truncate the number basis at some value of $n$ and $m$. The chosen value should be as high as possible, but it is important that it is at least larger than the mean photon number of the state. This will depend on the value of $\alpha_0$, and differ for coherent and squeezed states.
For example, the mean photon number of a coherent state relates to the amplitude as $\langle n \rangle = |\alpha|^2$, while for the squeezed state it also relates to the squeezing as $\langle n \rangle = |\alpha|^2 + \sinh^2 r$.

\subsubsection{Labelling}\label{sec:labelling}
In the paper~\cite{Havlicek2019}, labels were assigned according to the scheme: $\mathcal{L}(\tilde{x}) \geq \Delta \longrightarrow m(\tilde{x}) = +1$, and $\mathcal{L}(\tilde{x}) \leq -\Delta \longrightarrow m(\tilde{x}) = -1$, where $\Delta$ represents some gap that essentially cuts out data clustered close to zero. We saw no significant effect on learning compared to labels assigned based on points simply being above or below zero, so we  set $\Delta=0$. 

We assign labels according to:
\begin{align}
\mathcal{L}(\tilde{x}) &> 0 \longrightarrow m(\tilde{x}) = +1, \\
\text{and} ~\mathcal{L}(\tilde{x}) &< 0 \longrightarrow m(\tilde{x}) = -1.
\end{align}

\subsection{Training}\label{sec:training}
Data is divided into training and test sets. The algorithm learns on the training data, and then classifies the test data. As we have access to the labels for the test data, we will be able to verify the accuracy of the classification.

We use a support vector machine as our classifying method, specifically the SVM function in Python package scikit-learn, with our kernel defined as a ``custom kernel".

The kernel is defined\footnote{This particular example is for the coherent state with cross-Kerr nonlinearity}:

\begin{tcolorbox}
\begin{verbatim}
def kerr_theta(a,b,x1,x2,y1,y2):
  #Eq. 7 in this document
  theta = math.pi*((a**2)*(x1-y1)+(b**2)*(x2-y2)+2*a*b*((x1*x2)-(y1*y2)))
  return theta

def kerr_kernel(x,y):
  """Return value of kernel evaluated by ref state values of input variables"""
  ans = np.zeros((len(x),len(y)))
  print(np.shape(ans))
  for i in range(0,len(x)):
     for j in range(0,len(y)):
          x1 = x[i,0]
          x2 = x[i,1]
          y1 = y[j,0]
          y2 = y[j,1]
          ans[i][j] = np.exp(-2*abs(alpha)**2)*abs(sum([sum([np.exp(1j
          *kerr_theta(a,b,x1,x2,y1,y2))
          *((alpha**(2*(a+b)))/(math.factorial(a)*math.factorial(b)))
           for a in range(0,n)]) for b in range(0,m)]))**2
  return ans
\end{verbatim}
\end{tcolorbox}

This function is then called in the following code:
\begin{tcolorbox}
\begin{verbatim}
from sklearn import svm
h = .02  #Step size in the mesh
#Create an instance of SVM and fit data
kerr_svc = svm.SVC(kernel=kerr_kernel, C=***, gamma=***) #Input custom kernel
kerr_svc.fit(X_train, XZ_train) 
\
XZ_pred_kerr = kerr_svc.predict(X_test) #Predict response for test dataset
\end{verbatim}
\end{tcolorbox}

If not specified at the ``***'', hyperparameters default to $C=1.0$, and $\text{gamma}=$~``scale''.

\section{Results}

\subsection{Data Generation}\label{sec:datagen}
The initial data set of 100,000 points was generated in Mathematica from random values of $x_1, x_2$. The same raw data is then kept throughout the following analysis.

Some initial statistics for this raw data, presented in Table~\ref{tab:rawDataStats}, show that it is indeed randomly distributed between values on the unit square. 

\begin{table}[h]
\caption{Statistics for 100,000 data point set.}\label{tab:rawDataStats}
\centering
\begin{tabular}{lllllllllll}
 ~    & count   & mean     & std      & min       & 25\%      & 50\%     & 75\%     & max  & skew & kurtosis    \\
\hline
$x_1$    & 99999 & 0.499 & 0.289 & 0.000038  & 0.249  & 0.4982 & 0.7498 & 0.999995 & 0.00466 & -1.20037 \\
$x_2$    & 99999 & 0.500 & 0.289 & 0.000001  & 0.251  & 0.4995 & 0.7497 & 0.999999 & 0.00097 & -1.19714
\end{tabular}
\end{table}

\subsection{Coherent State Results}
The $x_1$ and $x_2$ values were then encoded into the coherent state with cross-Kerr nonlinearity of Equation~\eqref{eq:coherentencoding} plus the displaced parity operator.

The amplitude was set throughout to $\alpha = 1.0$. The number state basis is truncated at $n,m = 10$. This cutoff, combined with the value chosen for $\alpha$, is a trade off between keeping the values of $n$ and $m$ above the mean photon number, and what is computationally tractable.
\medskip

Four pairs of displacement operators $\mu, \nu$, listed in Table~\ref{tab:munutable}, were initially generated randomly from a distribution of Gaussian random complex variables,
\begin{equation}
    P(\mu,\nu) =   (2\pi\sigma)^{-1}e^{-(|\mu|^2+\nu|^2)/\sigma}.
\end{equation}
Note that when $\sigma=\bar{n}$ this is the Glauber-Sudarshan P-function for a thermal state where $\bar{n}$ is mean thermal excitation. 
The random displacements are generated with in Mathematica.

\begin{table}[h]
\caption{Random displacement coefficents used in the following analysis}\label{tab:munutable}
\centering
\begin{tabular}{lll}
Set Name & $\mu$ & $\nu$ \\
\hline
munu1    & $-0.468484 - 0.401083 i$ & $~0.64506 + 0.419369 i$ \\
munu2    & $-0.944138 + 1.154442 i$ & $-0.48727 - 0.971631 i$ \\
munu3    & $-0.734793 + 0.161473 i$ & $-1.02449 + 0.031192 i$ \\
munu4    & $~0.400004 + 0.603827 i$ & $-0.24941 + 0.765836 i$
\end{tabular}
\end{table}

\begin{figure}[htbp!]
\centering
\includegraphics[scale=0.5]{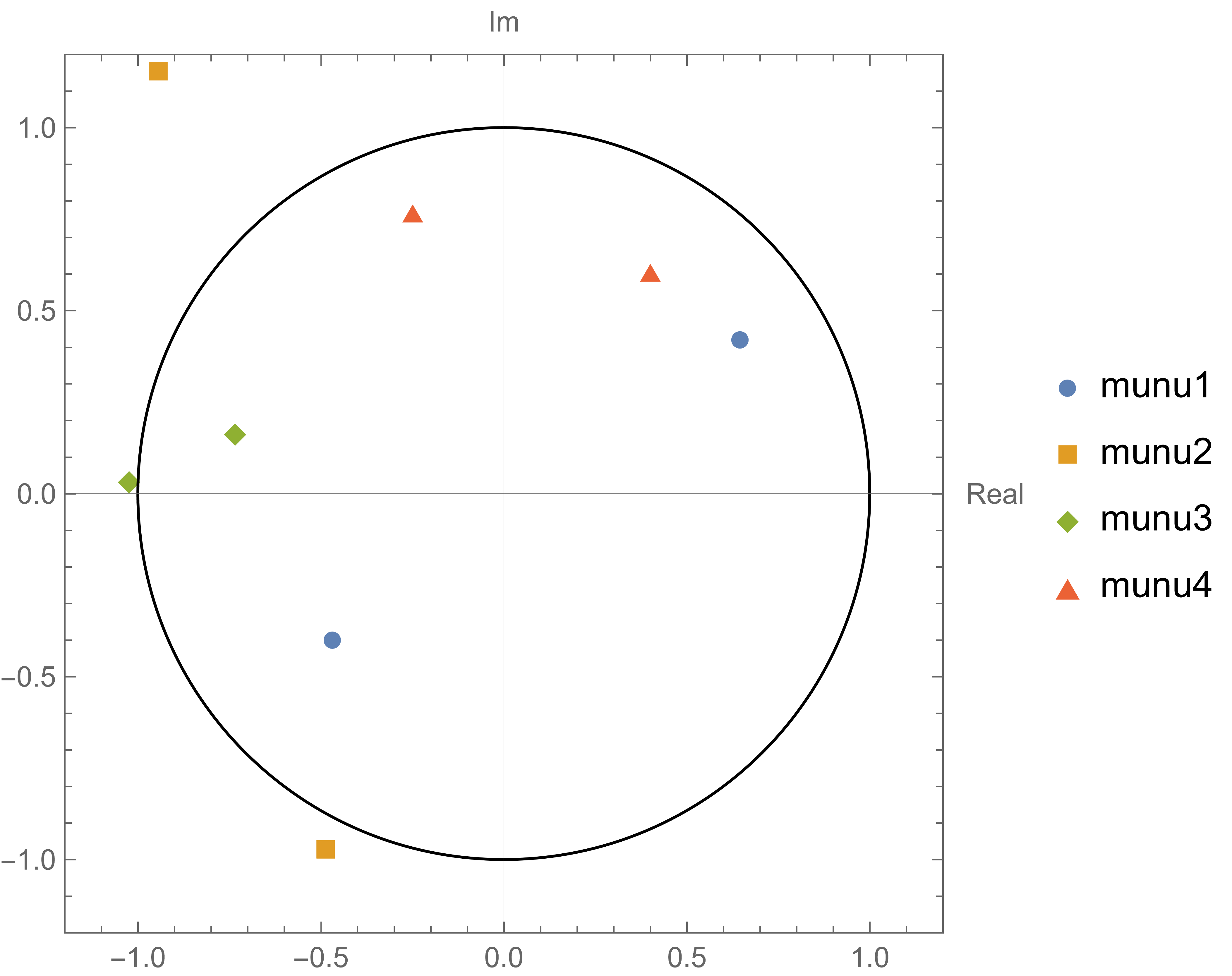}
 \caption{Distribution of displacement coefficients in the complex plane, drawn randomly from normal distributions with mean $0.0$ and variance $1.0$. Only values which fit criterion $|\mu|^2 < 3$ were used. As one can see, this did not ensure all were within the unit circle, but munu2 was kept to see what would happen.}\label{fig:complexplane}
\end{figure}

It is important that $|\mu|^2 ,|\nu|^2 < 3$ to keep the values inside the unit circle. This enables us to use a reasonable number state truncation to make the computation fast.  These eight displacement coefficients are plotted in Figure~\ref{fig:complexplane}, so one can see how they sit on the complex plane, and in relation to the circle.

\begin{figure}[htbp!]
\centering
\includegraphics[scale=0.5]{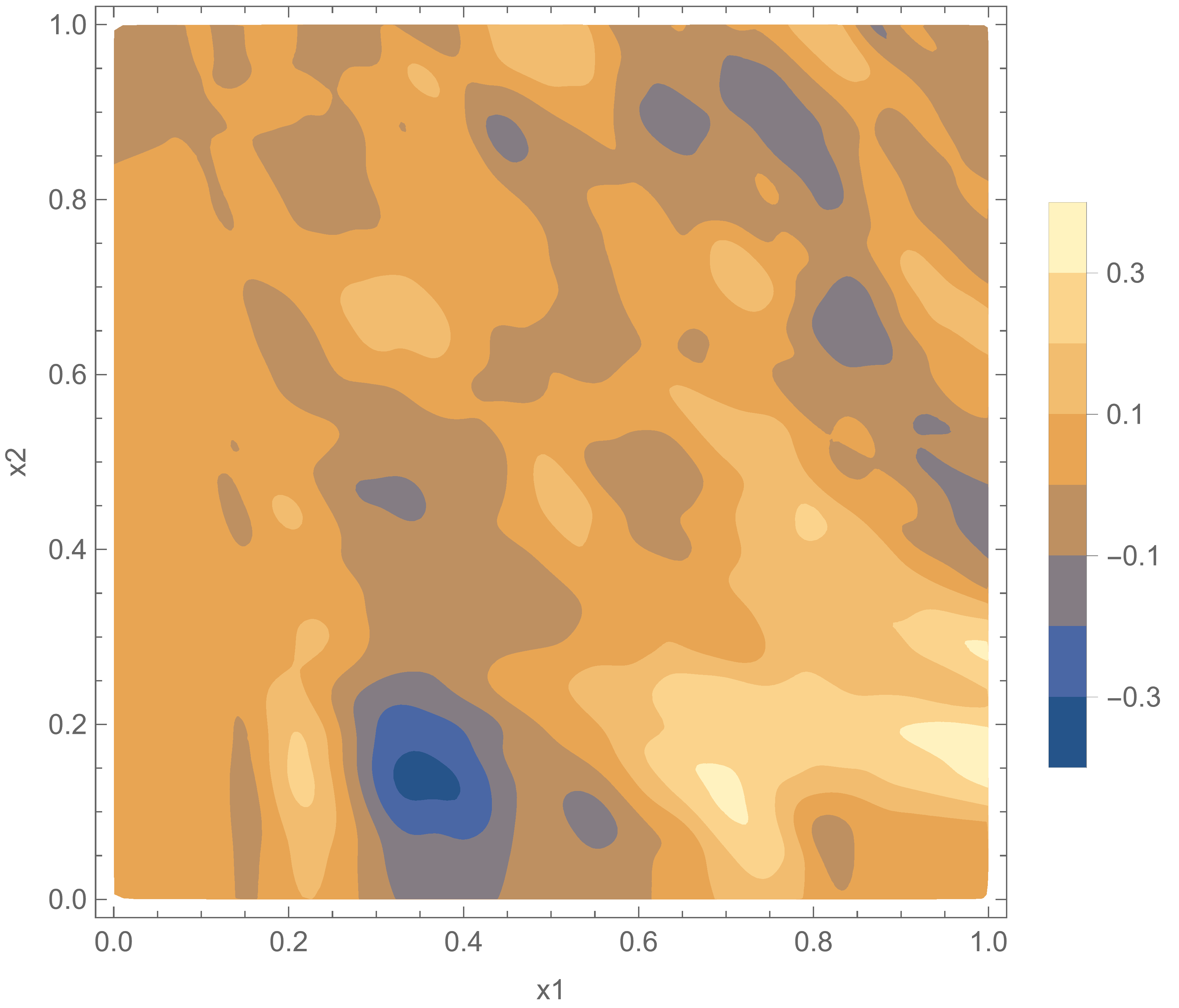}\includegraphics[scale=0.5]{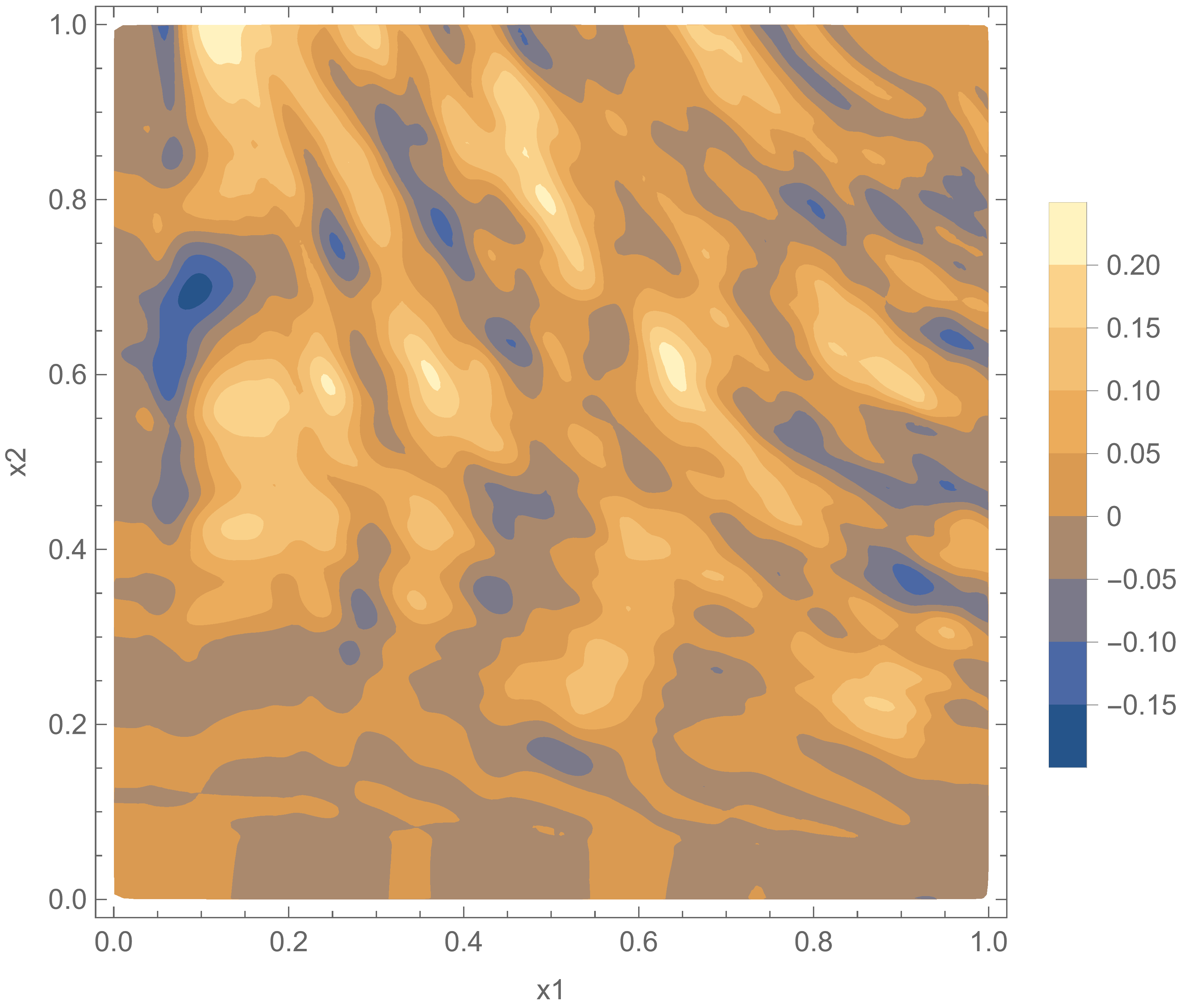}
\includegraphics[scale=0.5]{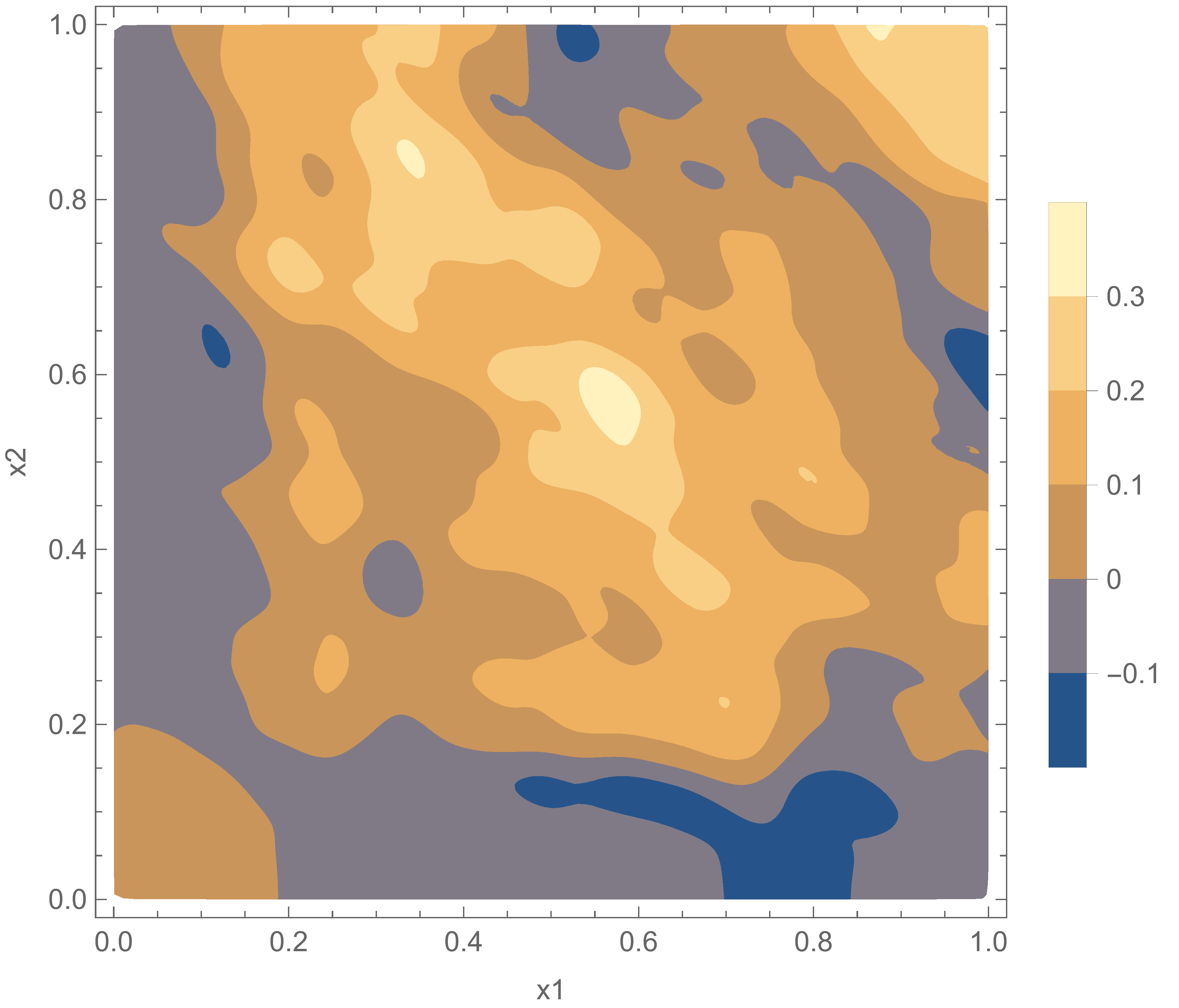}\includegraphics[scale=0.5]{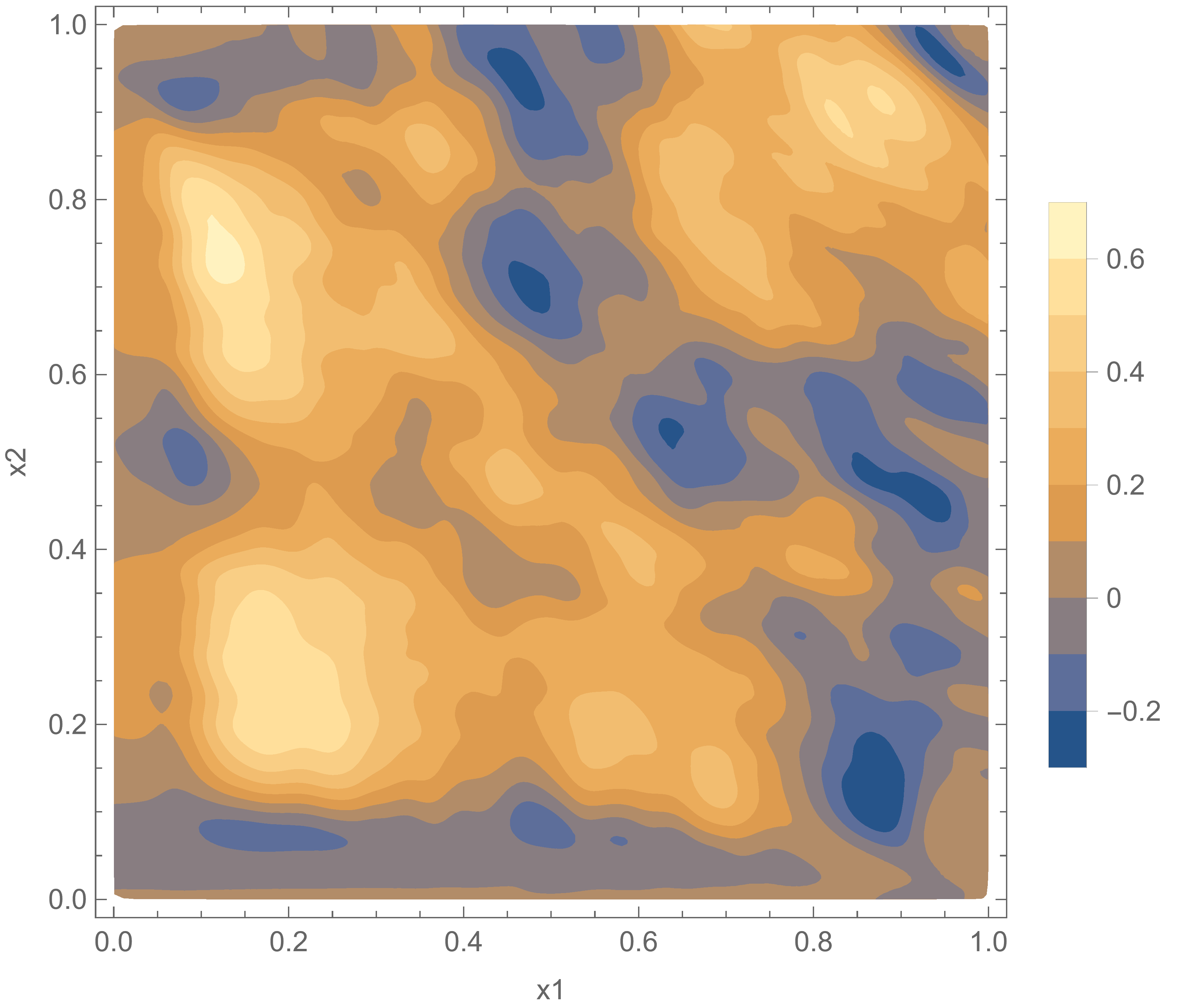}
  \caption{Contour plots of encoded data before any labelling. The top left figure is the `munu1' data, munu2 top right, munu3 bottom left, and munu4 bottom right (see Table~\ref{tab:munutable}).\label{fig:coherentmunucontours}}
\end{figure}

The resulting coherent-state-encoded data has the structure seen in Figure~\ref{fig:coherentmunucontours}. These are Wigner functions sampled at their respective values of $\mu$ and $\nu$.

\begin{figure}[htbp!]
\centering
\includegraphics[scale=0.5]{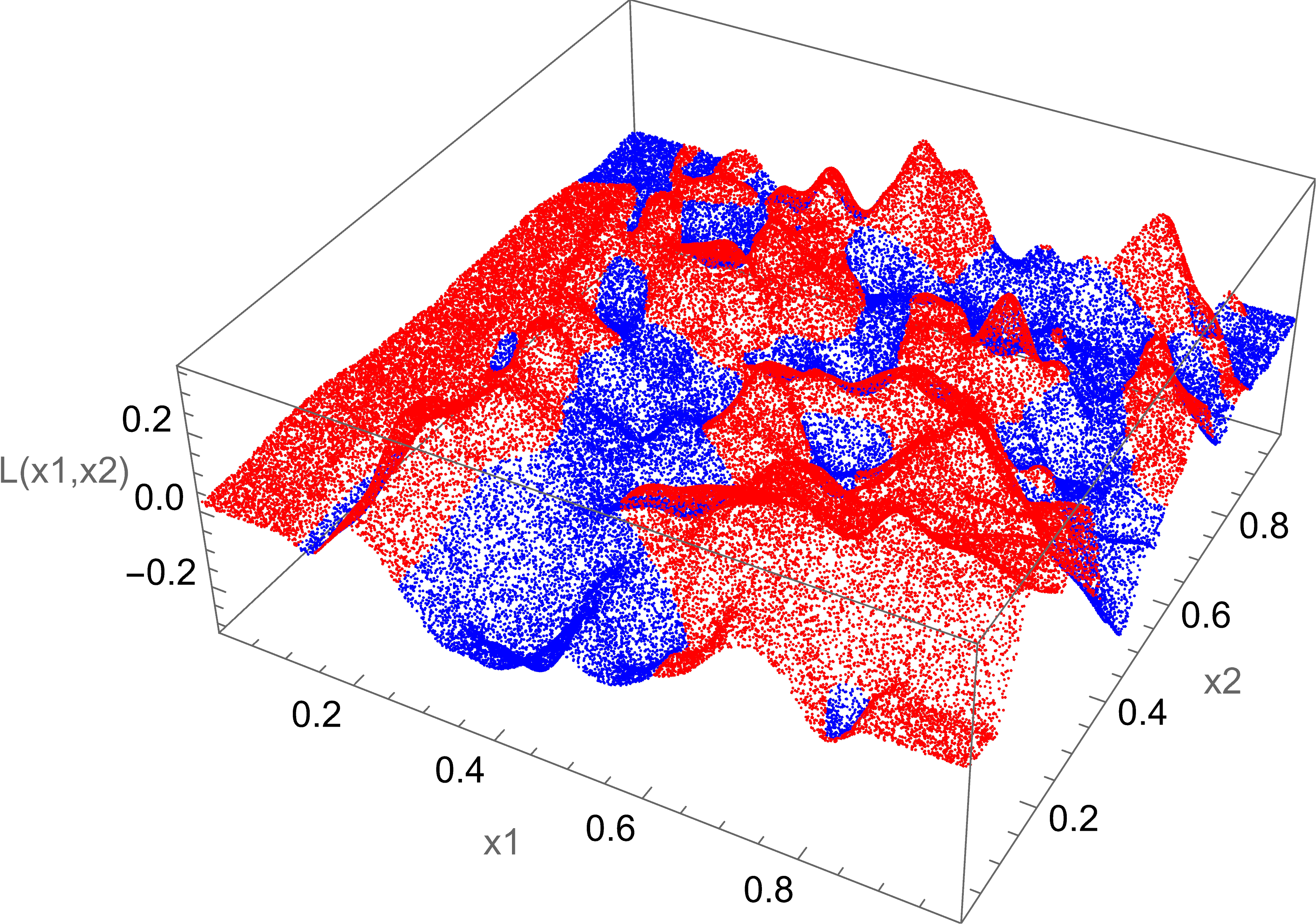}\includegraphics[scale=0.5]{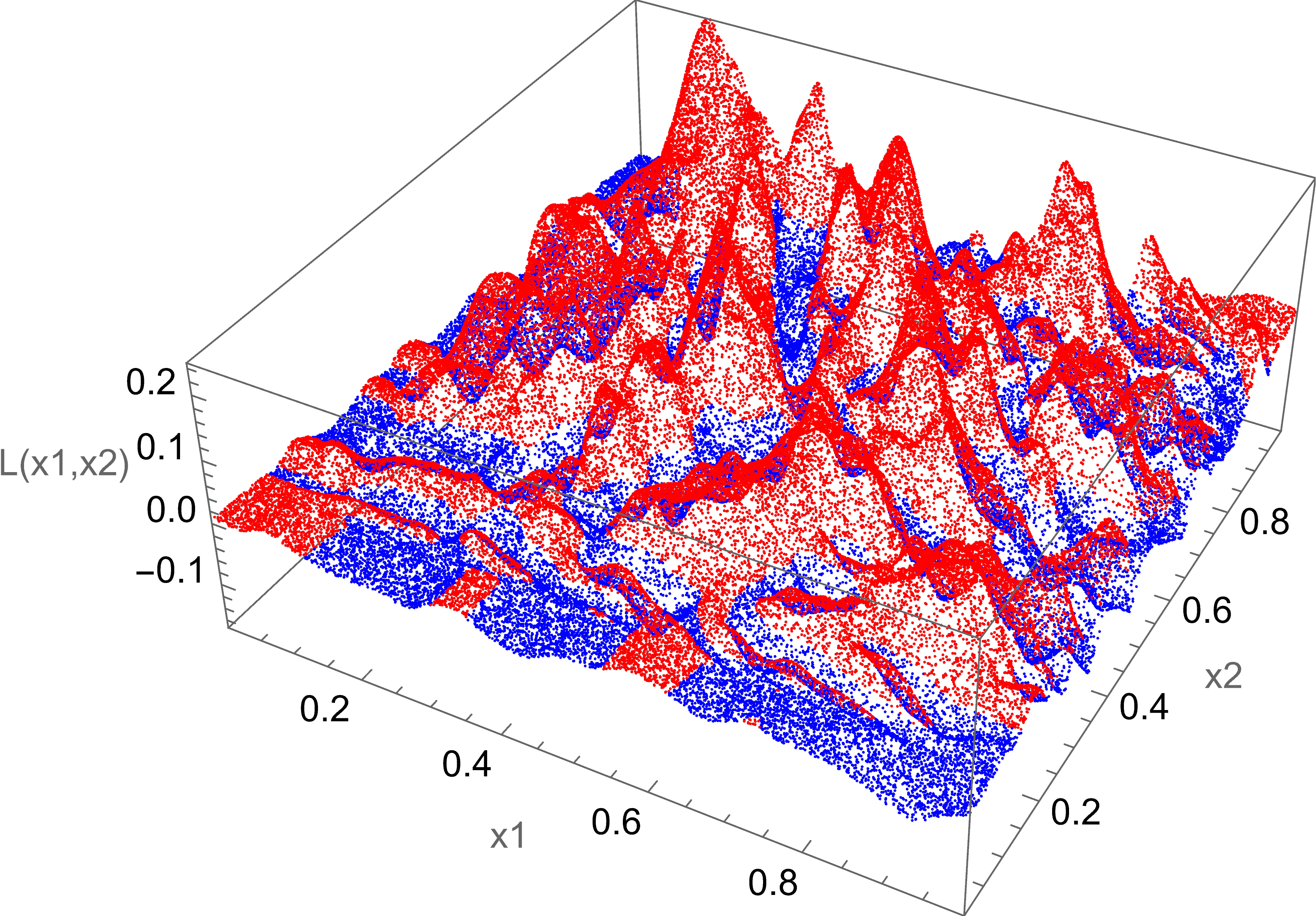}
\includegraphics[scale=0.5]{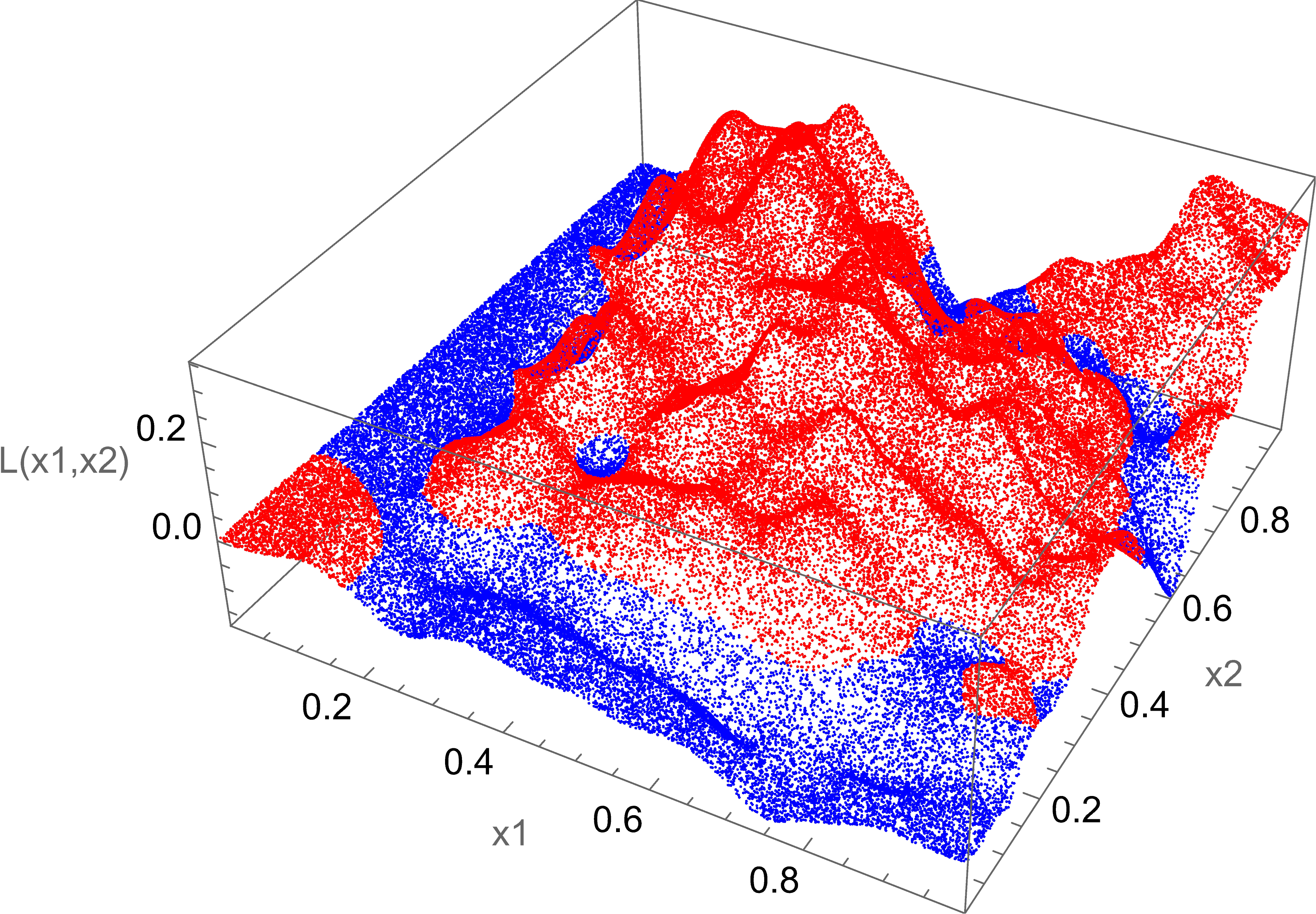}\includegraphics[scale=0.5]{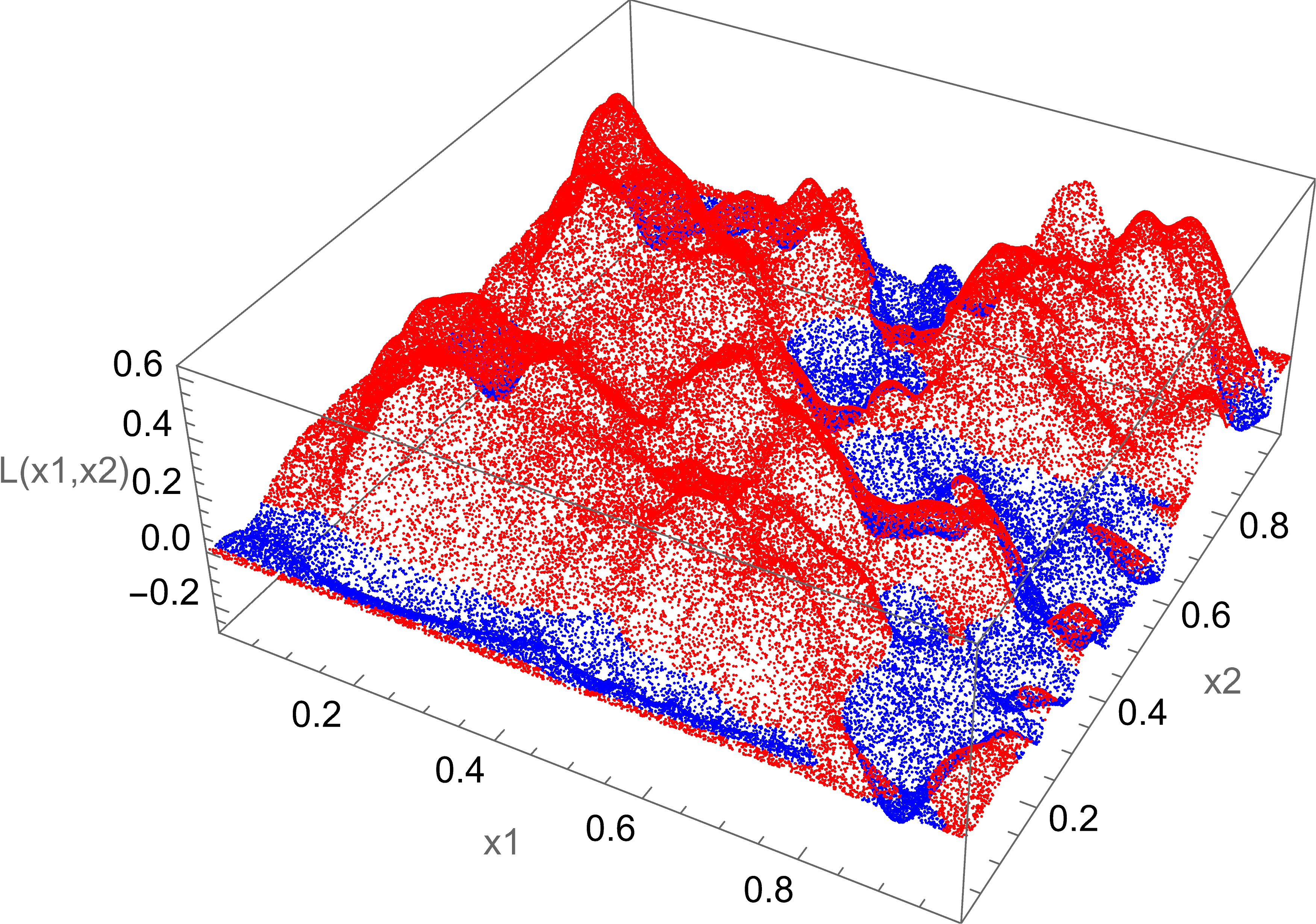}
  \caption{3D plots of labelled data. These are effectively Wigner functions measured at their respective values of $\mu$ and $\nu$. The top left figure is the `munu1' data, munu2 top right, munu3 bottom left, and munu4 bottom right (see Table~\ref{tab:munutable}).\label{fig:coherentmunuLabelledWigners}}
\end{figure}

The data is then labelled according to the scheme outlined in Section~\ref{sec:labelling}, with $\mathcal{L}(\tilde{x}) > 0$ assigning a $+1$ label and $\mathcal{L}(\tilde{x}) < 0$ assigning a $-1$. This results in different labelling for each of the four different combinations of displacement operators in Table~\ref{tab:munutable}. See the numbers per label in Table~\ref{tab:CoherentNumbers}, and Figure~\ref{fig:coherentmunuLabelledWigners} for the three-dimensional plots.

\begin{table}[h]
\caption{Numbers of data points with each label for the four data sets.}\label{tab:CoherentNumbers}
\centering
\begin{tabular}{llll|llll}
Set Name & Label & Label No. & Label Prop. & Set Name & Label & Label No. & Label Prop.\\
\hline
munu1    & $+1$  & 62304     & 0.623       & munu3    & $+1$  & 67856     & 0.679      \\
~        & $-1$  & 37695     & 0.377       & ~        & $-1$  & 32143     & 0.321      \\
munu2    & $+1$  & 59786     & 0.598       & munu4    & $+1$  & 71776     & 0.718      \\
~        & $-1$  & 40213     & 0.402       & ~        & $-1$  & 28223     & 0.282      \\
\end{tabular}
\end{table}

Data encoding and labelling was performed in Mathematica, we then move to Python for the SVM. 
\medskip

\begin{figure}[htbp!]
\centering
\includegraphics[scale=0.25]{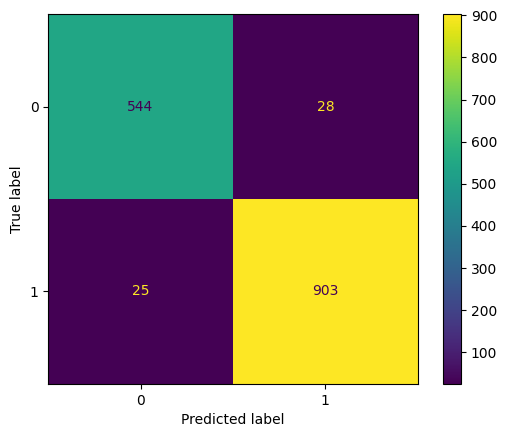}\includegraphics[scale=0.25]{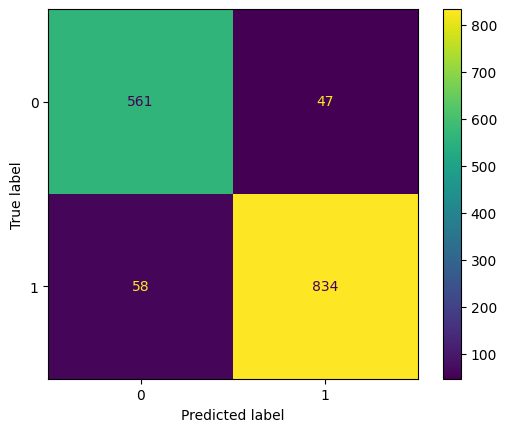}
\includegraphics[scale=0.25]{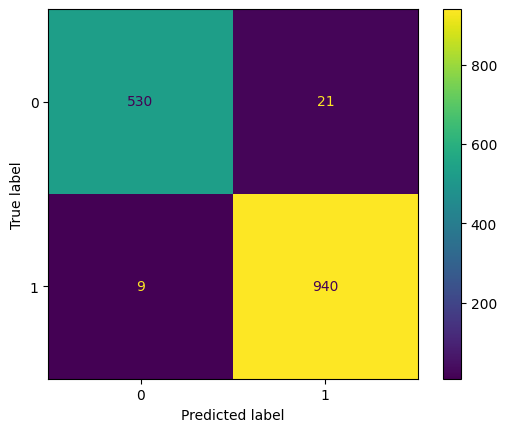}\includegraphics[scale=0.25]{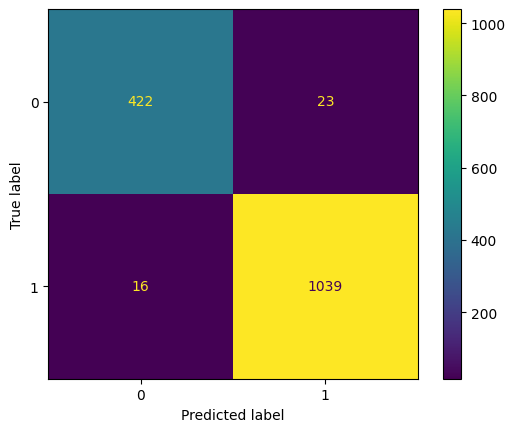}
  \caption{Confusion matrices for test data point labels vs predicted labels for each of the four data sets, with default hyperparameters. See Table~\ref{tab:coherentSVMresultsUntuned}. From initial sample of 5,000 data points, separated into sets of 3,500 for training and 1,500 for testing. Top left: munu1 data set, munu2 top right, munu3 bottom left, munu4 bottom right.\label{fig:coherentconfusionUntuned}}
\end{figure}

\begin{figure}[htbp!]
\centering
\includegraphics[scale=0.25]{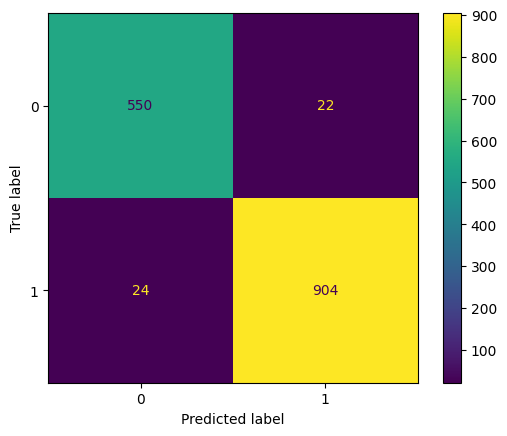}\includegraphics[scale=0.25]{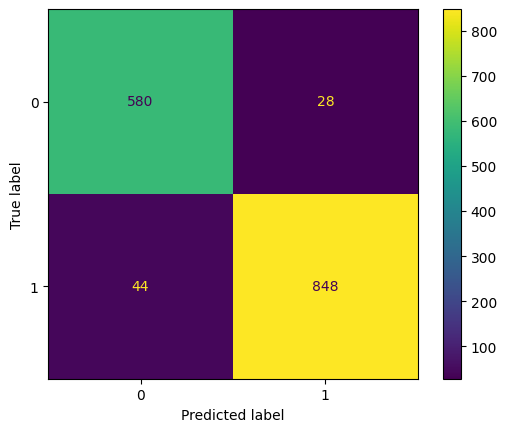}
\includegraphics[scale=0.25]{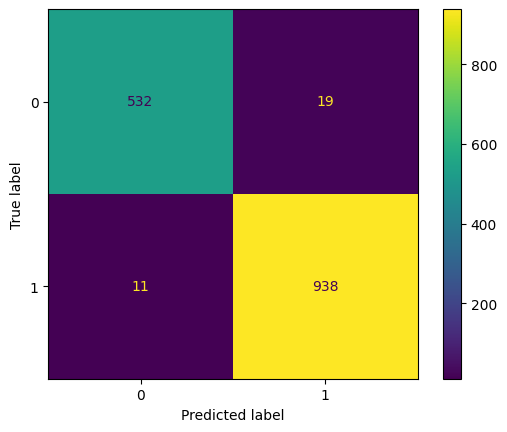}\includegraphics[scale=0.25]{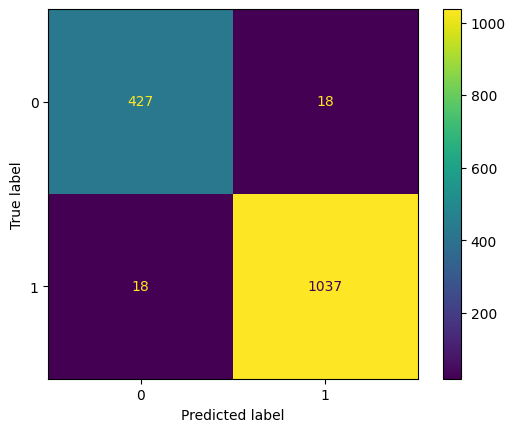}
  \caption{Confusion matrices for test data point labels vs predicted labels. From initial sample of 5,000 data points, separated into sets of 3,500 for training and 1,500 for testing. Values of C were determined via grid search, see Table~\ref{tab:coherentSVMresultsTuned}. Top left: munu1 data set, munu2 top right, munu3 bottom left, munu4 bottom right.\label{fig:coherentconfusionTuned}}
\end{figure}

We take samples of manageable size\footnote{Manageable for computation time - see tables for data set sizes.} from the first two columns of data, for each of the four data sets. That is, we take the raw $\vec{x}$ data points, not the encoded values. The sample is split into 70\% training and 30\% test data, and the scikit-learn SVM is run with our coherent Kerr kernel defined as a custom kernel as in the blocks of code given in Section~\ref{sec:training}. The results for the four data sets are given in Table~\ref{tab:coherentSVMresultsUntuned}, for default hyperparameters, and Table~\ref{tab:coherentSVMresultsTuned} for tuned values of C. The support vector machine's predicted labels for the test data are compared against the known (`true') labels in the confusion matrices in Figures~\ref{fig:coherentconfusionUntuned} and~\ref{fig:coherentconfusionTuned}, respectively.

\begin{figure}[htbp!]
\centering
\includegraphics[scale=0.5]{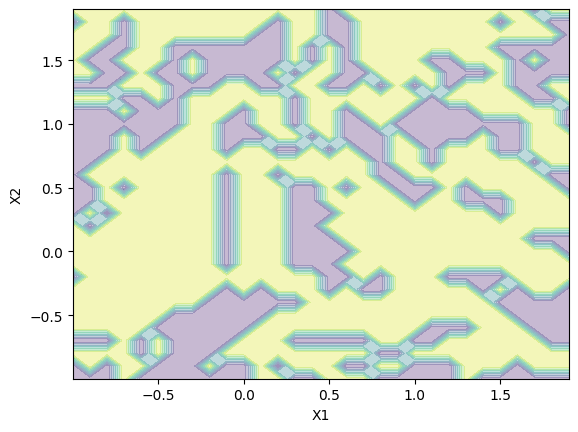}\includegraphics[scale=0.5]{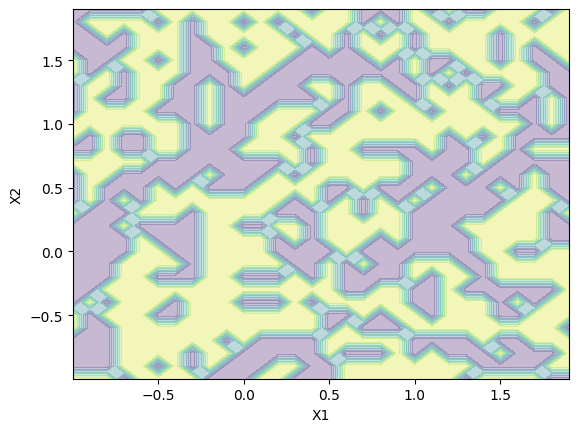}
\includegraphics[scale=0.5]{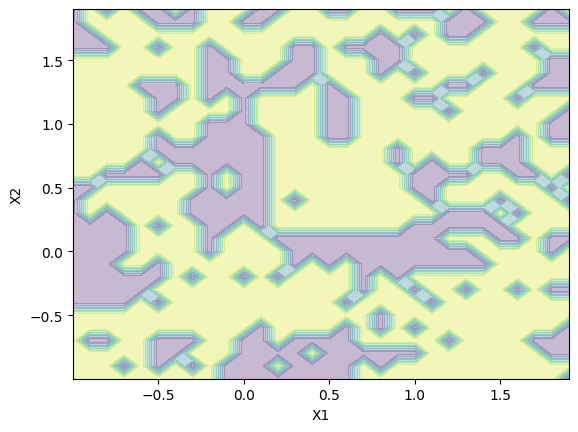}\includegraphics[scale=0.5]{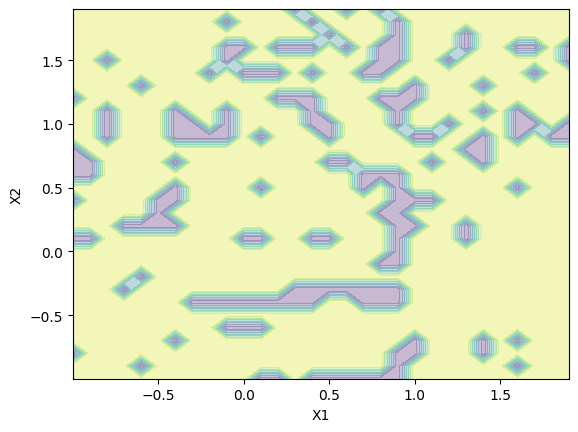}
  \caption{Decision boundaries determined by the support vector machine using our coherent Kerr kernel, for each of the four data sets, with default hyperparameters. See Table~\ref{tab:coherentSVMresultsUntuned}. From an initial sample of 5,000 data points, separated into sets of 3,500 for training and 1,500 for testing. Top left: munu1 data set, munu2 top right, munu3 bottom left, munu4 bottom right.\label{fig:coherentdecbound}}
\end{figure}

Decision boundaries for the untuned data are given in Figure~\ref{fig:coherentdecbound}.

\begin{table}[htbp!]
\caption{SVM results for the four data sets. Hyperparameters C and gamma were kept at the defaults 1.0 and ``scale", respectively. Corresponding confusion matrices in Figure~\ref{fig:coherentconfusionUntuned}, decision boundaries in Figure~\ref{fig:coherentdecbound}.\label{tab:coherentSVMresultsUntuned}}
\centering
\begin{tabular}{llllllll}
Set Name & Set Size & C & Gamma & Accuracy & Precision & Recall & f1-Score \\
\hline
munu1    & 5000     & 1 & scale & 0.9647   & 0.9699    & 0.9731 & 0.9689   \\
munu2    & 5000     & 1 & scale & 0.93     & 0.9467    & 0.9350 & 0.9408   \\
munu3    & 5000     & 1 & scale & 0.98     & 0.9781    & 0.9905 & 0.9843   \\
munu4    & 5000     & 1 & scale & 0.974    & 0.9783    & 0.9848 & 0.9815   \\
\end{tabular}
\end{table}

\begin{table}[htbp!]
\caption{SVM results for the four data sets. Hyperparameter C was set from grid search of different values of C. Gamma was kept at the default ``scale". Corresponding confusion matrices in Figure~\ref{fig:coherentconfusionTuned}\label{tab:coherentSVMresultsTuned}}
\centering
\begin{tabular}{llllllll}
Set Name & Set Size & C  & Gamma & Accuracy & Precision & Recall & f1-Score \\
\hline
munu1    & 5000     & 10 & scale & 0.9693   & 0.9762    & 0.9741 & 0.9751   \\
munu2    & 5000     & 10 & scale & 0.952    & 0.9680    & 0.9507 & 0.9593   \\
munu3    & 5000     & 10 & scale & 0.98     & 0.9801    & 0.9884 & 0.9842   \\
munu4    & 5000     & 10 & scale & 0.976    & 0.9829    & 0.9829 & 0.9829   \\
\end{tabular}
\end{table}

\subsubsection{Classical Kernel Comparison}\label{sec:coherentclassical}
The support vector machine was repeated with the radial basis function kernel. While the algorithm takes in only `raw' $x_1,x_2$ data, because labels are dependent on encoding, the analysis is still performed on each `munu' data set from Table~\ref{tab:munutable} in turn.
Default hyperparameters ($C=1.0$, gamma=scale) were used first, then tuned via grid search, see Table~\ref{tab:CoherentClassicalStats}.
\begin{table}[h]
\caption{SVM stats for coherent-state-encoded data sets as in Table~\ref{tab:munutable} using radial basis function (``classical'' kernel) for classification. Hyperparameters were the default $C=1.0$, gamma=scale first, and then both tuned via grid search.}\label{tab:CoherentClassicalStats}
\centering
\begin{tabular}{llllllll}
Set Name & Set Size & C     & Gamma   & Accuracy & Precision & Recall & f1-Score \\
\hline
munu1    & 5000     & 1     & scale & 0.8287   & 0.8372    & 0.8976 & 0.8663   \\
~        & 5000     & 1000  & 100   & 0.9533   & 0.9623    & 0.9623 & 0.9623   \\
\hline
munu2    & 5000     & 1     & scale & 0.7013   & 0.7044    & 0.8576 & 0.7735   \\
~        & 5000     & 100   & 1000  & 0.924    & 0.9502    & 0.9204 & 0.9351   \\
\hline
munu3    & 5000     & 1     & scale & 0.898    & 0.8734    & 0.9810 & 0.9241   \\
~        & 5000     & 1000  & 100   & 0.976    & 0.9731    & 0.9895 & 0.9812   \\
\hline
munu4    & 5000     & 1     & scale & 0.8213   & 0.8568    & 0.8957 & 0.8758   \\
~        & 5000     & 10000 & 100   & 0.9753   & 0.9820    & 0.9829 & 0.9824
\end{tabular}
\end{table}

\begin{figure}[h]
\centering
\includegraphics[scale=0.25]{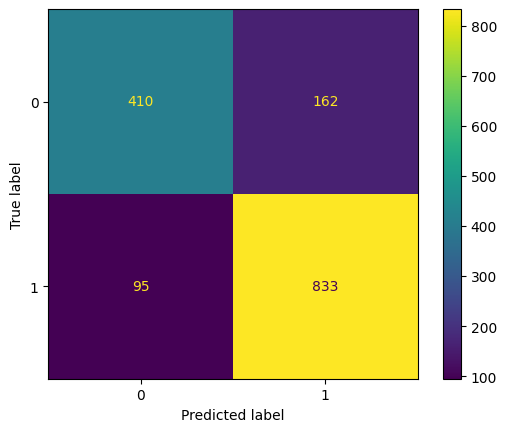}\includegraphics[scale=0.25]{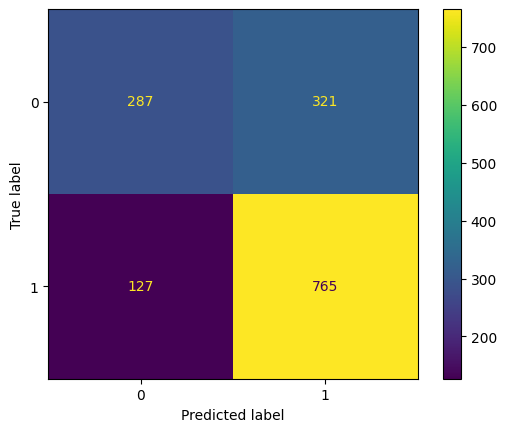}
\includegraphics[scale=0.25]{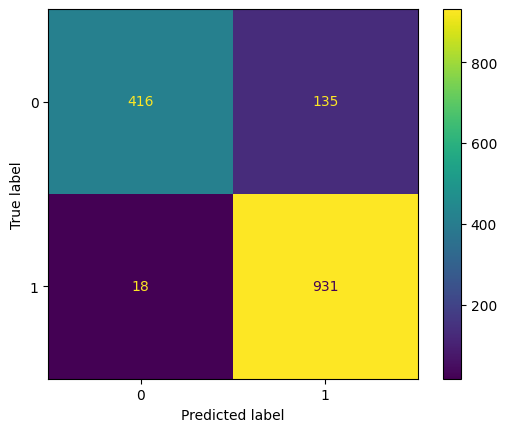}\includegraphics[scale=0.25]{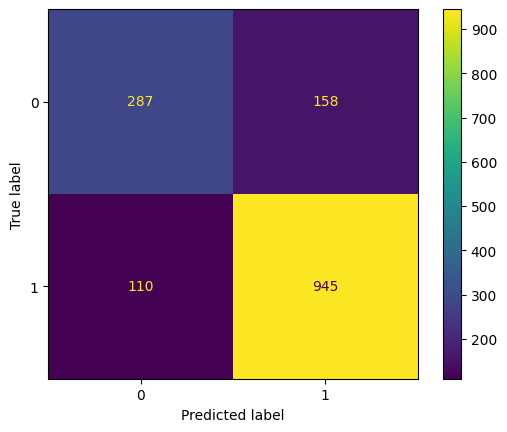}
  \caption{Confusion matrices for data sets in Table~\ref{tab:munutable} showing test data point labels vs predicted labels with the radial basis function kernel and default hyperparameters $C=1.0$, gamma=scale. Top left munu1 data set, top right munu2, bottom left munu3, bottom right munu4.}\label{fig:CoherentClassConf}
\end{figure}

\begin{figure}[h]
\centering
\includegraphics[scale=0.25]{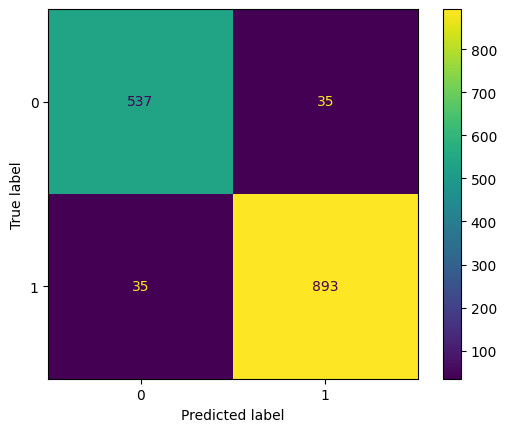}\includegraphics[scale=0.25]{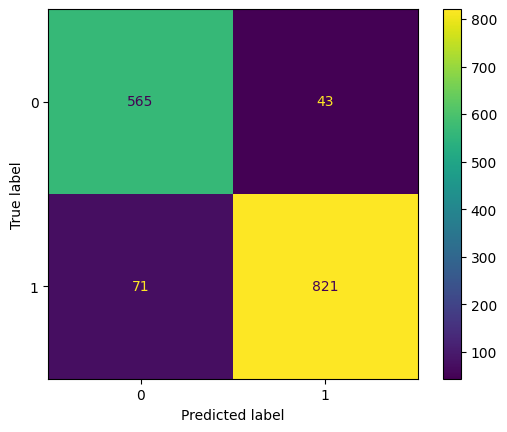}
\includegraphics[scale=0.25]{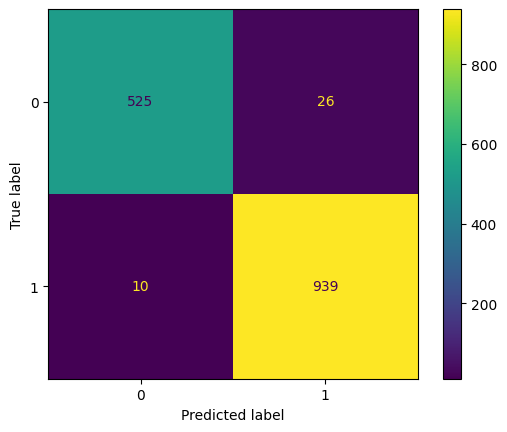}\includegraphics[scale=0.25]{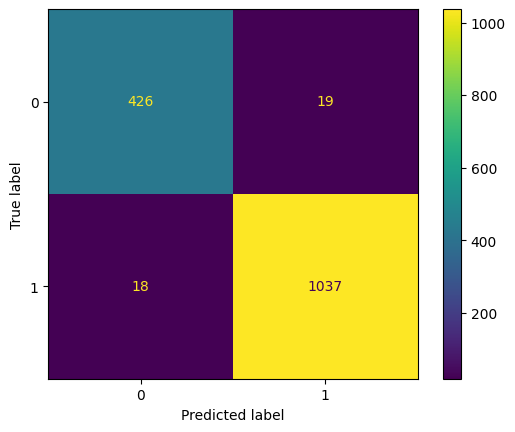}
  \caption{Confusion matrices for data sets in Table~\ref{tab:munutable} showing test data point labels vs predicted labels with the radial basis function kernel. Top left munu1 data set, top right munu2, bottom left munu3, bottom right munu4. Hyperparameters are tuned via grid search, as outlined in Table~\ref{tab:CoherentClassicalStats}.}\label{fig:CoherentClassConfTuned}
\end{figure}

\section{Effect of photon loss.}
As is well known,  cat states are highly sensitive to loss\cite{PhysRevA.31.2403}, so one might expect the quantum learning scheme proposed here will also require low loss. An initial density operator of the form
\begin{equation}
    \rho(0)={\cal N}(|\alpha_1\rangle\langle \alpha_1|+|\alpha_2\rangle\langle \alpha_2|+|\alpha_1\rangle\langle \alpha_2|+|\alpha_2\rangle\langle \alpha_1|
\end{equation}
undergoing spontaneous decay at rate $\gamma$ becomes
\begin{equation}
    \rho(t) ={\cal N}\sum_{i,j=1}^2\langle \alpha_i|\alpha_j\rangle ^{(1-e^{-\gamma t})}|\alpha_i e^{-\gamma t/2}\rangle\langle \alpha_j e^{-\gamma t/2}|
\end{equation}
For short times, 
\begin{equation}
  \left |\langle \alpha_i|\alpha_j\rangle\right | ^{2(1-e^{-\gamma t})}= e^{-\gamma t|\alpha_i-\alpha_j|^2}  
\end{equation}
This is for a cat state in which $\alpha_1=-\alpha_2=\alpha_0$ the rate of decay of the off diagonal term responsible for the negativity of the Wigner functions is $\gamma |\alpha_0|^2 $, much faster than the rate of amplitude decay, which is simply $\gamma$, if $|\alpha_0|^2 >> 1$. Thus photon loss will degrade the Wigner negativity required to distinguish the cats states in Fig. (\ref{new-WignerKC}), squandering the quantum advantage unless we use small cats, i.e. small values of $\alpha_0$.

In the single mode case, 
the dynamics including photon loss is described by the master equation
\begin{equation}
    \frac{d\rho}{dt}= -i\chi[(a^\dagger a)^2,\rho]+\gamma (a\rho a^\dagger -a^\dagger a\rho/2 -\rho a^\dagger a/2)
\end{equation}

For an initial coherent state, the solution, in the number basis, is \cite{HolMil}, 
\begin{equation}
    \rho_{nm}(t) =\rho_{nm}(0)f_{nm}(t)^{(n+m)/2}\exp\left [|\alpha_0|^2\frac{(1-f_{nm}(t)}{1+i\delta_{nm}}\right ]
\end{equation}
where
\begin{equation}
   \rho_{nm}(0) = e^{-|\alpha_0|^2}\frac{\alpha_0^n\alpha_0^{*\ m}}{\sqrt{n!m!}}
\end{equation}
and
\begin{eqnarray}
\delta_{nm}& = & 2\chi(n-m)/\gamma\\
    f_{nm}(t) & = &  e^{-\gamma  t-2i\chi t(n-m) }
\end{eqnarray}
The corresponding decision function is 
\begin{equation}
    d(\alpha,t) = \sum_{k=0}^\infty (-1)^k \sum_{n,m=0}^\infty d_{mk}(\alpha)d_{kn}(-\alpha) \rho_{nm}(t)
\end{equation}

\begin{figure}[htbp!]
    \centering
    \includegraphics[scale=0.5]{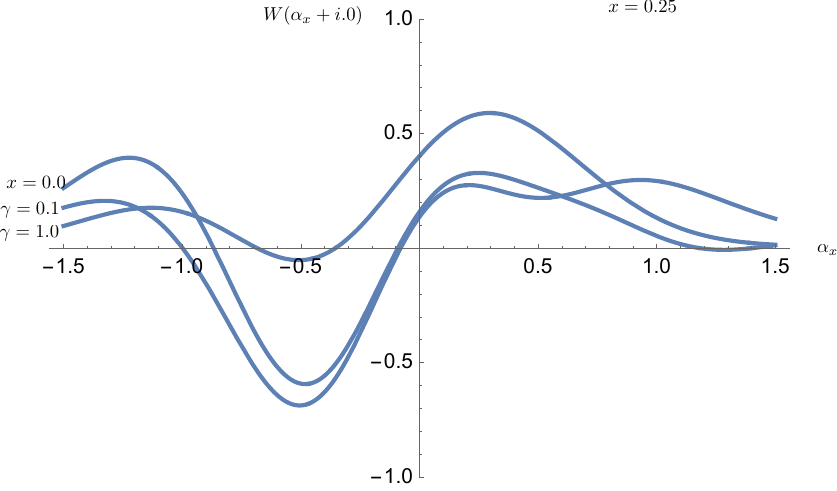}
    \caption{The Wigner function along the real axis for $|\phi(1/4)\rangle \ (x=1/4, \alpha_0=1.0) $ and different values of the dissipation rate $\gamma$.}
    \label{damped-wigner}
\end{figure}

The single mode case gives some insight into how loss affects the learning protocol. In the example of Fig. (\ref{new-WignerKC}) we see that the Wigner functions are distinguished along the real axis. The parameter update rule in Eq. (\ref{updates}) reflects this. We only need to look at this cross-section of the Wigner function.
In Fig.(\ref{damped-wigner}) we plot this cross-section Wigner function corresponding to $|\phi(1/4)\rangle$ for various values of $\gamma$. This should be compared to the unitary case in Fig. (\ref{wigner-minima}). There are two important features: significant regions of negativity remain for small $\gamma$ and the plateau is modulated into a single peak. 
The learning scheme we have proposed will continue to work for small values of loss so long as there is sufficient range  to distinguish the minima. The dependence on the loss rate gives an extra parameter to vary. While spontaneous emission is unavoidable it is easily included in bosonic system a controlled way using variable beam splitters. We suspect that learning can be optimised with respect to loss using this approach. One can also easily include phase space diffusion. This is described by a master equation of the form
\begin{equation}
    \dot{\rho}= {\cal L}\rho -\Gamma [\hat{n},[\hat{n},\rho]]
\end{equation}
where $\hat{n} = \sum_k a_k^\dagger a_k$\cite{WM}. This is easily implemented by adding a random phase to the choice of $\alpha_0$ in the fiducial state. This phase obeys a stochastic process with a phase diffusion  rate $\Gamma$. Adding this kind of noise may help overcome problems in quantum machine learning schemes such as barren plateaus and generalisation\cite{Eisert,bowles2024better}.


\section{Continuous time protocol.}

The sequential protocol lends it self to a continuous time implementation as discussed in \cite{CP-LM-review}. 
In this picture $x(t), y(t)$ are deterministic signals, perphaps a digital audio signal. However the output is a continuous stochastic process $d(t)$, with $|d(t)|\leq 1$, as are the displacement paramers, $\alpha(t)$. The learning process  leads to a complex stochastic process for $\alpha(t)$ that eventually makes $d(t)\approx y(t)$ in a statistical sense. The output signal is now constructed as a continuous measurement record derived from the discrete parity measurements. 

There is also a time continuous version of the parallel/kernel scenario. In this case, the kernel is a functional of two stochastic processes $x(t), y(t)$ written as $  K[x(t),y(t)]$ . This approach leads to the Karhunen-Lo\'{e}ve theory\cite{KL-theorem} for continuous stochastic processes. We will discuss this approach in a  future paper.

\section{Experimental implementation schemes.}
\label{exp}
Superconducting quantum circuits provide the ideal platform for Kerr non linear quantum dynamics\cite{Nori,Blais}. The Kerr effect can be 11 orders of magnitude larger than in nonlinear optical materials\cite{PhysRevB.98.094516}. The non linearity originates in the Josepheson effect and can be tuned using a variety of external controls. Quantum fields are typically confined by planar superconducting waveguides and measurement is done using direct detection of the electric field via heterodyne detection\cite{Krantz} and Josepheson parametric amplifiers\cite{Lehnert-JPA}.  While direct photon counting is difficult at microwave frequencies, there have been experimental demonstrations\cite{Opremcak}, including number resolving\cite{Curtis} detection.  

\begin{figure}
    \centering
    \includegraphics[scale=0.5]{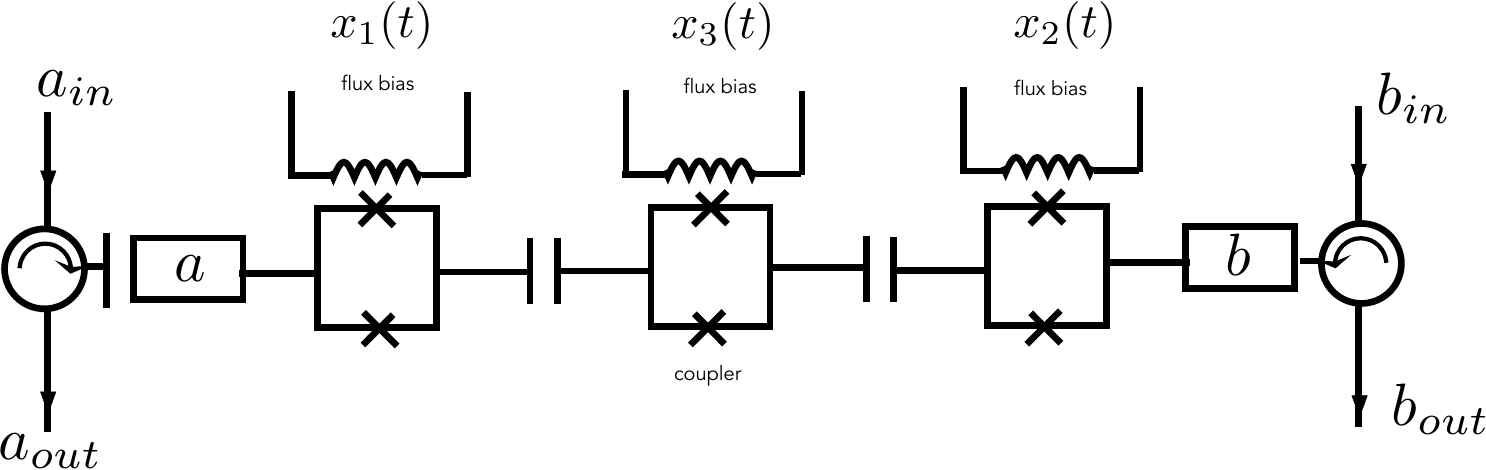}
    \caption{An idealised architecture to implement a two-mode quantum kernel evaluation. }
    \label{device-scheme}
\end{figure}

Entangled cat states in superconducting circuits were first demonstrated by Vlastakis et al. \cite{Vlastakis}.  The origin of the non linearity is the nonlinear inductance of a SQUID loop. The Kerr kernel described above can be implemented using the ability to fabricated tunable Kerr non linearities in coplanar resonators for microwave fields\cite{PhysRevA.105.023519}. The tuning is achieved by changing the flux bias of a DC SQUID. A tunable cross-Kerr interaction can also be implemented is such circuits\cite{Kounalakis}. The general scheme for a two-mode device is sketched in Fig. (\ref{device-scheme}). 

 \begin{figure}[htbp!]
     \centering
     \includegraphics[scale=0.5]{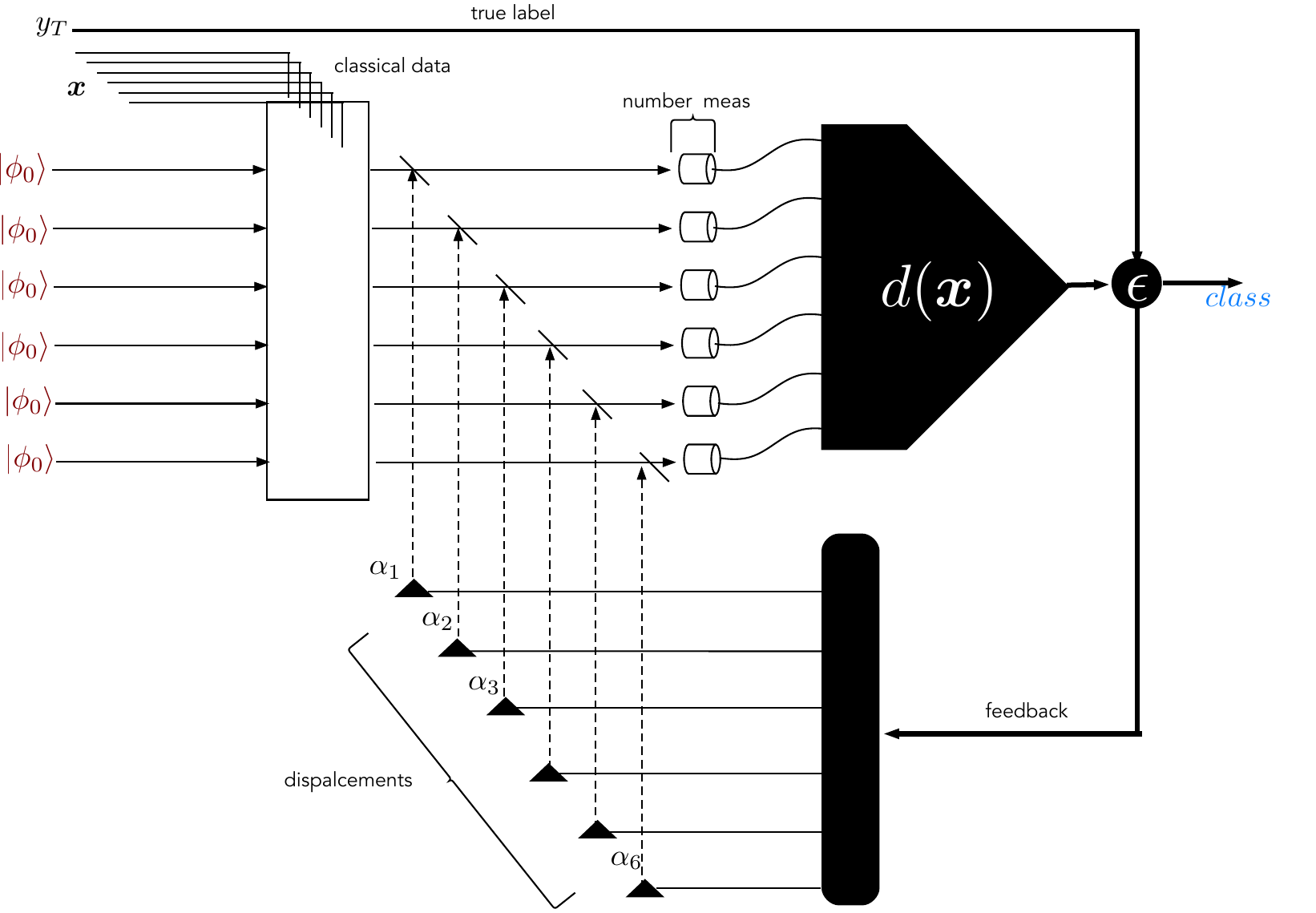}
     \caption{A schematic representation of a Kerr a bosonic quantum learning machine. The encoding is done using many modes interacting via cross Kerr non linear coupling. The data modulates the strength of the Kerr non linearity. The required displacements are implemented in the feedback loop using almost perfectly transmitting beam-splitters. Finally number resolving detectors are required to implement the feedback loop. }
     \label{multi-wigner}
 \end{figure}
The Wigner function for the single mode case is given by Eq. (\ref{alternating}). The function
\begin{equation}
 g(n,\mu,x)= \left |\langle \phi(x)|D(\mu)|n\rangle\right |^2
 \end{equation}
 is the probability to detect $n$ photons for the displaced state $D^\dagger(\mu)|\phi(x)\rangle$. Displacements are straightforward using the beam splitter interaction. The experiment requires  sampling this distribution. In our scheme the mean photon number is kept small by the choice of $\alpha_0$. This means the sampling size can be kept small.  
  We summarise the learning scheme in Fig. (\ref{multi-wigner}).

\section*{Acknowledgements}
We acknowledge support from the Australian Research Council Centre of Excellence for Engineered Quantum Systems (EQUS, CE170100009).

\newpage

\bibliography{Qkernel}

\end{document}